%% file: robust.tex
\documentclass[10pt,journal,compsoc]{IEEEtran}
\ifCLASSOPTIONcompsoc
  \usepackage[noadjust]{cite}
  
\else
  \usepackage{cite}
\fi
\ifCLASSINFOpdf
  \usepackage[pdftex]{graphicx}
\else
  \usepackage[dvips]{graphicx}
\fi
\usepackage{amsmath}
\usepackage{amssymb}
\usepackage{pifont}

\usepackage[normalem]{ulem}

\usepackage{algorithm}
\usepackage{algorithmicx}
\usepackage{algpseudocode}
\usepackage{mathrsfs}
\usepackage{amsthm}

\usepackage{wrapfig}
\usepackage{diagbox}

\usepackage{url}

\usepackage{color}
\usepackage{multirow}
\usepackage{stackrel}
\usepackage{amsopn}

\theoremstyle{plain}

\algnewcommand\algorithmicforeach{\textbf{for each:}}
\algnewcommand\ForEach{\item[ \algorithmicforeach]}

\begin{document}
%
\title{FTK: A Simplicial Spacetime Meshing Framework for Robust and Scalable\\Feature Tracking}
%
%
%
%

\author{Hanqi~Guo,~\IEEEmembership{Member,~IEEE,}
        David~Lenz,
        Jiayi~Xu,
        Xin~Liang,
        Wenbin~He,
        Iulian~R.~Grindeanu,
        Han-Wei~Shen,~\IEEEmembership{Member,~IEEE,}
        Tom~Peterka,~\IEEEmembership{Member,~IEEE,}
        Todd~Munson,
        and~Ian~Foster~\IEEEmembership{Fellow,~IEEE}
\IEEEcompsocitemizethanks{\IEEEcompsocthanksitem Hanqi Guo is with the Mathematics and Computer Science Division, Argonne National Laboratory, Lemont, IL 60439, USA.\protect\\
E-mail: hguo@anl.gov
\IEEEcompsocthanksitem David Lenz is with the Mathematics and Computer Science Division, Argonne National Laboratory, Lemont, IL 60439, USA.\protect\\
E-mail: dlenz@anl.gov
\IEEEcompsocthanksitem Jiayi Xu is with the Department of Computer Science and Engineering, the Ohio State University, Columbus, OH 43210, USA.\protect\\
E-mail: xu.2205@osu.edu
\IEEEcompsocthanksitem Xin Liang is with the Department of Computer Science,  Missouri University of Science and Technology, Rolla, MO 65409, USA.\protect\\
E-mail: xliang@mst.edu
\IEEEcompsocthanksitem Iulian R. Grindeanu is with the Mathematics and Computer Science Division, Argonne National Laboratory, Lemont, IL 60439, USA.\protect\\
E-mail: iulian@mcs.anl.gov
\IEEEcompsocthanksitem Wenbin He is with Bosch Research North America, Sunnyvale, CA 94085, USA.\protect\\
E-mail: wenbin.he2@us.bosch.com
\IEEEcompsocthanksitem Han-Wei Shen is with the Department of Computer Science and Engineering, the Ohio State University, Columbus, OH 43210, USA.\protect\\
E-mail: shen.94@osu.edu
\IEEEcompsocthanksitem Tom Peterka is with the Mathematics and Computer Science Division, Argonne National Laboratory, Lemont, IL 60439, USA.\protect\\
E-mail: tpeterka@mcs.anl.gov
\IEEEcompsocthanksitem Todd Munson is with the Mathematics and Computer Science Division, Argonne National Laboratory, Lemont, IL 60439, USA.\protect\\
E-mail: tmunson@mcs.anl.gov
\IEEEcompsocthanksitem Ian Foster is with the Data Science and Learning Division, Argonne National Laboratory, Lemont, IL 60439, USA.\protect\\
E-mail: foster@anl.gov}
}

%
%

\markboth{IEEE Transactions of Visualization and Computer Graphics,~Vol.~X, No.~X, November~2021}%
{Guo \MakeLowercase{\textit{et al.}}: A Simplicial Spacetime Meshing Framework for Robust and Scalable Feature Tracking}
%



\IEEEtitleabstractindextext{%
\begin{abstract}
We present the Feature Tracking Kit (FTK), a framework that simplifies, scales, and delivers various feature-tracking algorithms for scientific data.  The key of FTK is our simplicial spacetime meshing scheme that generalizes both regular and unstructured spatial meshes to spacetime while tessellating spacetime mesh elements into simplices.  The benefits of using simplicial spacetime meshes include (1) reducing ambiguity cases for feature extraction and tracking, (2) simplifying the handling of degeneracies using symbolic perturbations, and (3) enabling scalable and parallel processing.  The use of simplicial spacetime meshing simplifies and improves the implementation of several feature-tracking algorithms for critical points, quantum vortices, and isosurfaces.  As a software framework, FTK provides end users with VTK/ParaView filters, Python bindings, a command line interface, and programming interfaces for feature-tracking applications.  We demonstrate use cases as well as scalability studies through both synthetic data and scientific applications including tokamak, fluid dynamics, and superconductivity simulations.  We also conduct end-to-end performance studies on the Summit supercomputer.  FTK is open sourced under the MIT license: \url{https://github.com/hguo/ftk}.
\end{abstract}
%
\begin{IEEEkeywords}
Feature tracking, spacetime meshing, distributed and parallel processing, critical points, isosurfaces, vortices.
\end{IEEEkeywords}}

\maketitle

\IEEEdisplaynontitleabstractindextext

%

\allowdisplaybreaks

\input{introduction.tex}
\input{background.tex}
\input{mesh.tex}
\input{critical.tex}
\input{vortex.tex}

\input{contour.tex}

\input{library.tex}

\input{interface.tex}
\input{performance.tex}
\input{discussion.tex}
\input{conclusions.tex}
\ifCLASSOPTIONcompsoc
  \section*{Acknowledgments}
\else
  \section*{Acknowledgment}
\fi

We thank reviewers for their  valuable feedback.  This research is supported by the Exascale Computing Project (ECP), project number 17-SC-20-SC, a collaborative effort of the U.S. Department of Energy Office of Science and the National Nuclear Security Administration, as part of the Co-design center for Online Data Analysis and Reduction (CODAR)~\cite{CODAR2020}.  It is also supported by the U.S. Department of Energy, Office of Advanced Scientific Computing Research, Scientific Discovery through Advanced Computing (SciDAC) program, and by Laboratory Directed Research and Development (LDRD) funding from Argonne National Laboratory, provided by the Director, Office of Science, of the U.S. Department of Energy under Contract No. DE-AC02-06CH11357.
This work is also supported in part by National Science Foundation Division of Information and Intelligent Systems-1955764.

\ifCLASSOPTIONcaptionsoff
\fi


\bibliographystyle{IEEEtran}
\bibliography{IEEEabrv,robust}
%

%





{\footnotesize The submitted manuscript has been created by UChicago Argonne, LLC, Operator of Argonne National Laboratory (``Argonne''). Argonne, a U.S. Department of Energy Office of Science laboratory, is operated under Contract No. DE-AC02-06CH11357. The U.S. Government retains for itself, and others acting on its behalf, a paid-up, nonexclusive, irrevocable worldwide license in said article to reproduce, prepare derivative works, distribute copies to the public, and perform publicly and display publicly, by or on behalf of the Government.The Department of Energy will provide public access to these results of federally sponsored research in accordance with the DOE Public Access Plan. http://energy.gov/downloads/doe-public-access-plan.}

\vfill



\end{document}

%% file: introduction.tex
\section{Introduction}

Feature tracking is a core topic in scientific visualization for  understanding  dynamic behaviors in time-varying simulation and experimental data.  By tracking features such as extrema, vortex cores, and boundary surfaces, one can  highlight key regions in visualization,  reduce data to store, and  enable further analysis based on the dynamics of features in scientific data.

This paper introduces a general framework that delivers a collection of feature-tracking tools to end users, scales feature-tracking algorithms in distributed and parallel environments, and simplifies the development of new feature-tracking algorithms.  The motivations for developing this framework are threefold.  
First, although feature-tracking capabilities appear sporadically in today's data analysis and visualization tools,  a general-purpose toolset is lacking that would enable users to track and analyze features in scientific workflows.  In community tools such as VTK~\cite{vtk}, VTK-m~\cite{MorelandSULMPKS16}, ParaView~\cite{paraview}, VisIt~\cite{VisIt}, and TTK~\cite{TiernyFLGM18}, most algorithms focus on single-timestep data, and only a few filters are provided for tracking features over time.  Object tracking for videos is available in computer vision libraries such as OpenCV~\cite{opencv}, but scientific data differ significantly from natural videos in their feature definitions and data representation.  
Second, few existing feature-tracking algorithms are designed for scalability and parallel processing.  The advent of exascale computing means that data produced by supercomputers need to be efficiently handled by the same scale of computing resources.  In both in situ and post hoc scenarios, the sheer data size and high complexity of tracking algorithms necessitate distributing data to many computing nodes and using GPU accelerators when available. 
Third, no developer framework exists for  eliminating redundant efforts to implement application-specific feature-tracking algorithms.  Implementing feature tracking algorithms from scratch can be daunting; the management of time-varying data, the handling of degenerate cases, and the parallelization of tracking algorithms are needed in many applications; such features do not exist in publicly available software libraries.

To these ends, we identify the common ground---spacetime meshing---among many tracking algorithms for isosurfaces, critical points, and vortex cores.  By extruding the spatial mesh into the time dimension, a spacetime mesh connects the cells in the spatial mesh over adjacent timesteps.  For example, in 3D isosurface tracking, marching cubes~\cite{LorensenC87} are generalized to higher dimensions~\cite{BhaniramkaWC04, JiSW03} by iterating and classifying 4D spacetime cells with lookup tables.  In critical point tracking~\cite{TricocheWSH02, GarthTS04}, the movement of critical points can be captured by identifying the spatiotemporal cells that contain critical points.  Likewise, in tracking quantum vortices in complex-valued scalar fields~\cite{GuoPPKG16, GuoPG17}, the moving trajectories of vortex core lines can be reconstructed in spacetime meshes.

We present the Feature Tracking Kit (FTK), which introduces \emph{simplicial spacetime meshing} for robust and scalable feature tracking.  Compared with spacetime meshes used previously, the key difference of our method is that all mesh elements are \emph{simplices}.  Previous feature-tracking methods extruded 2D triangles into 3D prisms~\cite{TricocheWSH02, GuoPG17} and 3D cubes into 4D cubes~\cite{BhaniramkaWC04, GuoPPKG16}; but neither a prism nor a cube in the extruded mesh is simplicial.  

Simplicial meshes offer three benefits for feature tracking: specificity, stability, and scalability.  
First, simplicial meshes eliminate ambiguities in feature tracking, similar to how marching tetrahedra~\cite{DoiK91} eliminates isosurface ambiguity.  In nonsimplicial cells such as cubes, multiple features intersect the same cell, causing ambiguities that require attention.  We show that with the spacetime piecewise linearity (PL) assumption, no disambiguation is needed for tracking critical points and isosurfaces in simplicial meshes.  
Second, simplicial meshes ease the handling of degeneracies, enabling robust feature tracking.  Degeneracies, such as a critical point on an edge or isosurface intersecting a vertex, may lead to loss or duplication of the detection results due to numerical instabilities~\cite{BhatiaGWBP14}.  Simplicial meshes enable the use of Simulation of Simplicity (SoS)~\cite{EdelsbrunnerM94a}---a mature programming technique to simplify the handling of degeneracies in computational geometry---to generate robust, combinatorial, and consistent tracking results regardless of numerical instabilities.  
Third, simplicial meshes make it straightforward to accelerate feature-tracking algorithms with both GPU parallelism and distributed parallelism.  In cases when the feature detection is independent in each cell, we can distribute the tasks to different computing resources for concurrent and scalable processing.

In this study, we design and implement the simplicial subdivision of two types of spacetime meshes---$(n+1)$-D prismatic and $(n+1)$-D regular meshes---to enable robust and scalable feature tracking in both unstructured and regular meshes, in order to support the tracking of critical points (0D features in 2D/3D), quantum vortices (1D features in 3D), and isosurfaces (2D features in 3D) for a wide range of applications.  The $(n+1)$-dimensional space consists of both 2D/3D space and time, and all mesh elements are simplices.  Enabled by simplicial spacetime meshing, each individual tracking algorithm has novelties in disambiguation, degeneracy handling, and scalability.  
We also enable efficient mesh element traversal over time. Considering that time-varying data are large and streamed from simulations in situ, one can iterate spacetime mesh elements within a sliding window of a few timesteps for out-of-core and streaming data access.  

As a software framework, FTK provides ParaView plugins, Python bindings, command line interfaces, and programming interfaces for end users to track a variety of features both in situ and post hoc.  We demonstrate the use of FTK for fluid dynamics, fusion, and superconductivity simulations.
In summary, the novelty of this paper is in its combination of several individual technical contributions:  
\begin{itemize}
  \item A simplicial spacetime meshing scheme that generalizes and tessellates both regular and unstructured spatial meshes to spacetime simplices (Section~\ref{sec:hypermesh}) 
  \item A robust and scalable critical point tracking algorithm that handles degeneracies in a consistent manner with no ambiguities (Section~\ref{sec:critical})
  \item A scalable implementation of quantum vortex tracking with distributed parallelism (Section~\ref{sec:vortex})
  \item A robust and scalable isosurface tracking algorithm that avoids ambiguities and handles degeneracies in a consistent manner (Section~\ref{sec:contour})
  \item A software framework for users to track features with distributed and parallel environments both in situ and post hoc (Sections~\ref{sec:library} and \ref{sec:interface})
  \item Comprehensive performance studies of FTK algorithms on both supercomputers and commodity hardware (Section~\ref{sec:performance})
\end{itemize}

%% file: background.tex
\section{Related Work}
\label{section:background}

This section reviews related work in the extraction and tracking of critical points, quantum vortices, and isosurfaces.  We also briefly review simulation of simplicity and spacetime meshing.  In the following, we refer to  the process of independently detecting features in individual timesteps as feature extraction, as opposed to feature tracking, which is the process of reconstructing trajectories of features through multiple consecutive timesteps.  
For a comprehensive review of feature extraction and tracking, see Post et al.~\cite{PostVHLD03}; 
 for a review of topology-based methods for visualization, see Heine et al.~\cite{HeineLHIFSHG16}.  


\subsection{Critical point extraction and tracking}

In general, a critical point is defined as the location where a vector field vanishes.  
This work treats critical points in scalar fields through their gradient fields; limitations of this treatment will be discussed in later sections.
Below, we review critical point extraction and tracking algorithms in both vector and scalar fields.

\subsubsection{Critical point extraction}

\textbf{Numerical methods} have been used to locate critical points where all vector components are zero simultaneously, assuming the vector field can be interpolated based on discrete representations.  While finding such zero crossings 
has been studied for bilinear and trilinear schemes~\cite{HaynesP07}, the piecewise linear (PL) interpolation of vector fields is more widely used in various applications because of its simplicity~\cite{HelmanH89, TricocheSH00, LiangGDCRLOCP20}.  Tricoche et al.~\cite{TricocheSH00} characterized higher-order critical points in 2D PL vector fields by partitioning neighboring regions based on different flow behavior.  That approach was generalized to 3D vector fields~\cite{WeinkaufTSHS05}.  Besides PL, extraction of critical points in piecewise-constant vector fields can be achieved by discrete Hodge decomposition~\cite{PolthierP02}. 

\begin{figure}[h]
  \includegraphics[width=\linewidth]{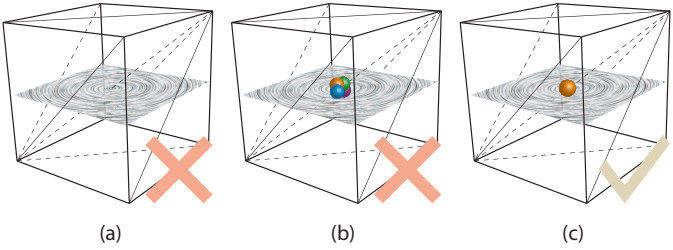}
  \caption{Nonrobust (a and b) versus robust (c) critical point extraction.  With numerical methods, when a critical point resides on an edge, the critical point may or may not be detected by all tetrahedron that share the same edge; in this case, the number of detected critical points could range from zero to six because of numerical instabilities. 
  With the help of Simulation of Simplicities~\cite{EdelsbrunnerM94a, BhatiaGWBP14}, the single critical point will be detected and associated with one of the triangles in a combinatorial manner.}
  \label{fig:robust}
\end{figure}

A major issue with numerical methods is their sensitivity to numerical instabilities.  As illustrated in Figures~\ref{fig:robust}(a) and (b), a critical point may be identified multiple times if the critical point resides on the boundary of cells.  To this end, Bhatia et al.~\cite{BhatiaGWBP14} introduced the use of Simulation of Simplices (SoS)~\cite{EdelsbrunnerM94a} in testing whether a simplicial cell contains critical points, leading to a robust critical point detection in a combinatorial manner, illustrated in Figure~\ref{fig:robust}(c).  Our study further generalizes the use of SoS to ensure that the tracking of critical points is robust and combinatorial, as demonstrated in following sections.  

In scalar fields, numerical methods for extracting critical points have been studied in the context of resolving ambiguities of marching cubes.  In the early 1990s, Nielson and Hamann~\cite{NielsonH91} derived the closed-form representation of critical points (saddles) in bilinear interpolants, in order to handle ambiguous quadrilateral faces.  Nielson~\cite{Nielson03}  generalized the derivation to trilinear interpolants, which can be used to subdivide an ambiguous cube into blocks with simple configurations~\cite{CarrM09}.  Carr and Snoeyink~\cite{CarrS09} further studied scalar field topology through contour trees that capture bilinear/trilinear interpolant topologies.


\textbf{Topology methods} include the use of the Poincar\'e index theorem, (discrete) Morse theories, and contour trees/Reeb graphs.   For example, Poincar\'e's index theorem can be used to test whether critical points exist in 3D regions~\cite{Greene92, MannR02} or PL surfaces~\cite{LiVRL06}.  With Morse decomposition, Chen et al.~\cite{ChenMLZ08} proposed a vector field topology representation of 2D PL vector fields with graphs, such that critical points can be identified as part of the vector field topology.  For scalar fields, critical points are the key constituents of the scalar field topology, including Reeb graphs~\cite{Reeb46} and contour trees~\cite{CarrSA00} extracted with well-established algorithms.



\subsubsection{Critical point tracking}

\textbf{Spacetime meshing methods} perform interpolation over time to track critical points.  Tricoche et al.~\cite{TricocheWSH02} extruded 2D triangular cells into 3D spacetime prisms, detected entries and exits of singularities on prism faces, and then identified paths of critical points.  Garth et al.~\cite{GarthTS04} generalized this approach to 3D by extruding from tetrahedra to 4D tetrahedral prisms.  For both methods, the vector field is assumed linear over both space (barycentric interpolation) and time (linear interpolation).  As with bilinear and trilinear interpolations, such spacetime interpolation in prisms is not PL in spacetime, as explained in the next section.  


\textbf{Feature flow field (FFF) methods}~\cite{TheiselS03} use a derived FFF vector field to characterize feature movements, such that feature trajectories can be computed as tangent curves in FFF.  For critical point tracking, one needs to find an appropriate set of critical points in spacetime as the seeds, compute tangent curves from the seeds by numerical integration, and then slice the tangent curves back into individual timesteps.  To address instabilities in the numerical integration, Weinkauf et al.~\cite{WeinkaufTGP11} proposed a method to improve convergence.  Klein and Ertl generalized FFF to track critical points in scale space.~\cite{KleinE07}  Reininghaus et al.~\cite{ReininghausKWH12} proposed a combinatorial version of FFF based on the discrete Morse theory to track critical points in 2D scalar fields.

\textbf{Nearest-neighbor and region-overlapping approaches} are heuristics to track critical points.  For example, Wang et al.~\cite{WangRSBP13} reconstructed critical point trajectories by joining proximal and same-type critical points in adjacent timesteps.  Skraba and Wang~\cite{SkrabaW14} used the closeness of robustness, the minimum amount of perturbation needed to cancel features, to link corresponding critical points in adjacent timesteps. 



\subsection{Quantum vortex extraction and tracking}

We use quantum vortices as an example of tracking 1D features.  Quantum vortices, or simply vortices, are topological defects in superconductivity~\cite{PhillipsPKG15}, superfluidity~\cite{GuoLXXF18}, and Bose--Einstein condensates.  Simulations produce 3D complex-valued fields that combine both amplitudes and phase angles.  Singularities in phase fields are closed 1D curves embedded in 3D Euclidean spaces.  By definition, a vortex is  the locus of points such that 
\begin{equation}\label{eq:vortex}
  -\oint_C \nabla \theta(\mathbf{x}) \cdot d\mathbf{l} = 2k\pi, k\neq0 ,
\end{equation}
where $\theta(\mathbf{x})$ is the phase field, $C$ is an infinitesimal contour that encircles the vortex curve, $d\mathbf{l}$ is the infinitesimal arc on $C$, and $k$ is a nonzero integer usually equal to $\pm1$.  

\textbf{Vortex extraction.}\quad Based on the definition, a straightforward approach to extract vortices in 3D meshes is to first check whether the contour integral is nonzero for each face boundary and then to trace singularity curves along faces~\cite{PhillipsPKG15, GuoPPKG16}.  
Guo et al.~\cite{GuoPG17} proved that a triangulated mesh cell intersects up to one singularity line and thus that simplicial mesh subdivision leads to combinatorial and consistent extraction results.  

\textbf{Vortex tracking.}\quad A spacetime meshing approach was proposed to associate vortex curves in adjacent timesteps~\cite{GuoPPKG16, PhillipsGPKG16}. 
As a result, mesh faces testing positive for a singularity form graphs that characterize the movement of singularities as surfaces.  Guo et al.~\cite{GuoPG17} used triangular/tetrahedral prisms  as the spacetime cells to extract and track singularities.  However, ambiguities still exist because spacetime prisms are nonsimplicial.  In Section~\ref{sec:vortex} we demonstrate that a simplicial spacetime mesh eliminates ambiguities in a consistent manner and allows parallel vortex curve tracking in distributed environments. 

Quantum vortices are fundamentally different from vortices in fluid flows~\cite{JiangMT05}, which are swirling centers of flows and have been defined by level sets or extremum lines of $\lambda_2$ eigenvalues~\cite{JeongH95} and vorticity magnitude~\cite{ZabuskyBPGSC91}.  Depending on definitions, tracking of fluid flow vortices may be achieved by connected component labeling~\cite{SilverW98} or FFF~\cite{TheiselS03}.

\subsection{Isosurface extraction and tracking}

\textbf{Isosurface extraction}---the task of reconstructing polygon surfaces with a given isovalue in a 3D scalar field---is fundamental to scientific visualization.  The marching cubes~\cite{LorensenC87} algorithm extracts isosurfaces in regular grid data based on lookup tables; disambiguating how surfaces are connected inside a cube was the key research problem for a decade~\cite{NielsonH91, Nielson03, CarrM09, CarrS09}.  A cubic cell has $2^8=256$ possible ways to intersect an isosurface, which boil down to 15 unique configurations.  Ambiguities exist when vertex values have alternating signs on any faces.  Marching tetrahedra~\cite{DoiK91} is a promising method to eliminate ambiguities by tessellating inputs into simplicial cells; each tetrahedron has  only two unambiguous cases of intersections.  For rectilinear grid data, different approaches exist for subdividing cubes into simplices~\cite{CarrMS06}, which may induce visual artifacts and topology variations.

\textbf{Isosurface tracking.}\quad Two distinct approaches exist for tracking isosurfaces: volume tracking~\cite{SilverW96, SilverW98} and higher-dimensional isosurfacing~\cite{BhaniramkaWC04}.  The former extracts regions of interest independently in each timestep and then associates regions across adjacent timesteps based on overlaps.  The latter generalizes marching cubes to 4D spacetime; the outputs are 3D objects embedded in 4D that can be sliced back into 2D surfaces in individual timesteps for visualization.  Similar to marching cubes in 3D, ambiguities exist in 4D spacetime, and researchers have shown that one can disambiguate 4D cases by triangulation~\cite{JiSW03}.  We will demonstrate in Section~\ref{sec:contour} a simplified implementation of 4D isosurface tracking based on our simplicial spacetime meshes.

\subsection{Simulation of Simplicity}
\label{sec:sos}

Simulation of Simplicity (SoS)~\cite{EdelsbrunnerM94a} is a programming technique to simplify the handling of degenerate cases in geometric algorithms.  In 1D spaces, an analogy to SoS is the  stable sorting algorithm, which handles degenerate cases when the input array contains equal numbers; if two numbers are equal, the number with the smaller array index is considered smaller than the other, resolving in consistent ordering of numbers.

\begin{figure}[h]
  \includegraphics[width=\linewidth]{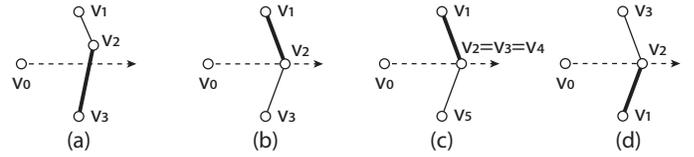}
  \caption{Example use of SoS for testing whether an edge intersects the horizontal half-line originated from $v_0$: (a) edge $v_2v_3$ intersects the half-line with no ambiguity; (b) ambiguity exists because $v_2$ intersects the half-line, and SoS rules that edge $v_1v_2$ is intersected while $v_2v_3$ is not; (c) with multiple overlapped vertices ($v_2$, $v_3$, and $v_4$) intersecting the half-line, SoS rules that edge $v_1v_2$ is intersected while edges $v_2v_3$, $v_3v_4$, and $v_4v_5$ are not; (d) by flipping the indices of $v_1$ and $v_3$ in (b), SoS resolves the ambiguity in a different way.}
  \label{fig:sos}
\end{figure}

We use the example of a 2D point-in-polygon test to explain the idea of SoS.  The test is positive if the count of intersections between edges and the horizontal half-line started with the test point ($v_0$ in Figure~\ref{fig:sos}(a)) is an odd number.  The half-line/edge intersection is tested by the sign of a determinant, and the test is unambiguous in nondegenerate cases.  Ambiguity occurs when the line intersects a vertex (Figure~\ref{fig:sos}(b)), because the determinant is zero.  By incorporating vertex indices, SoS simplifies the test by implicitly associating the intersection to one of the edges in a consistent manner, even in more complicated cases when vertices overlap (Figure~\ref{fig:sos}(c)).  Note that results may change if a different vertex ordering system is used (Figure~\ref{fig:sos}(d)), but this is normally not a problem because the ordering is predetermined.  FTK uses SoS to simplify the test if a simplex (e.g., edge, triangular face, or tetrahedron cell) intersects a feature (e.g., critical point trajectory or contour) in spacetime, as discussed in following sections.

\subsection{Spacetime meshing}

\textbf{Spacetime meshing for computational sciences.}\quad Recently, scientists have started to explore the use of 4D meshes~\cite{IshiiFSKGS19, Thite09} to numerically solve time-dependent partial differential equations in spacetime as opposed to traditional timestepping approaches.  We believe that our method could be directly applied to spacetime mesh data; but because of challenges of increased complexity, memory footprint, and cost to converge, the majority of scientific data today is still stored and represented in discrete timesteps.

\textbf{Spacetime meshing for scientific visualization.}\quad Spacetime meshing approaches, which are limited mostly to prisms to date, have been successfully used to track singularities in vector fields~\cite{TricocheWSH02, GarthTS04} and phase fields~\cite{PhillipsPKG15, GuoPPKG16, PhillipsGPKG16, GuoPG17}.  Prisms are a straightforward choice, but challenges exist in handling ambiguities, as discussed in later sections.

%% file: mesh.tex
\section{Simplicial Spacetime Mesh}
\label{sec:hypermesh}

We design and implement the simplicial subdivision of $(n+1)$-D prismatic and $(n+1)$-D regular meshes, respectively, in order to enable robust and scalable feature tracking in unstructured and regular grid meshes, where the dimensionality $n$ is 2 or 3 for the spatial domain.  The additional dimension is time in this study, and we assume that the spatial mesh does not change over time.  The simplicial subdivision of $(n+1)$-D regular meshes, which is a special case of the subdivision of $(n+1)$-D prismatic meshes, is implemented separately for the efficient handling of images, volumes, and curvilinear grids.  

\begin{figure*}
  \includegraphics[width=\linewidth]{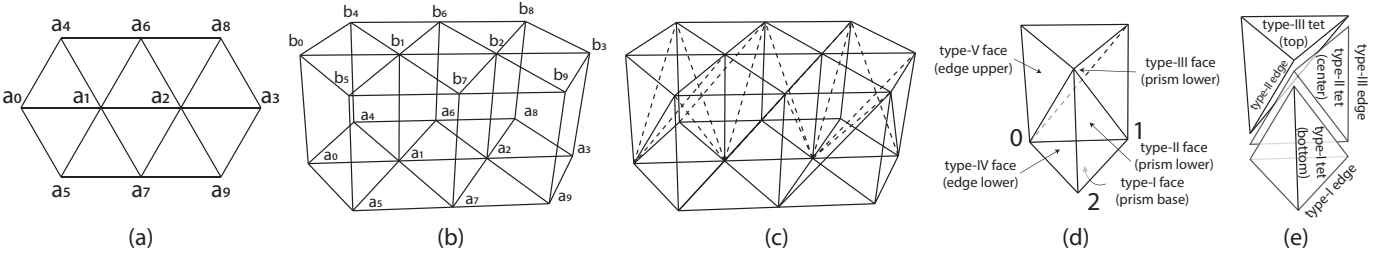}
  \caption{Extrusion of simplicial mesh: (a) the input 2D simplicial mesh, (b) the extruded 3D prismatic mesh, (c) the output 3D simplicial mesh as the subdivision of the 3D prismatic mesh, and (d and e) subdivision of a 3D triangular prism.  Unique types of edges, faces, and tetrahedra in the extruded mesh are illustrated in d and e.}
  \label{fig:triangular}
\end{figure*}

For example, in the case of $n=2$, the input is a (time-invariant) triangular mesh (illustrated in Figure~\ref{fig:triangular}(a)).  One can extrude the mesh into 3D by replicating and elevating vertices in the new dimension, forming 3D triangular prisms (Figure~\ref{fig:triangular}(b)).  The output 3D mesh is a subdivision of triangular prisms, and each mesh element in the output mesh is simplicial (Figure~\ref{fig:triangular}(c)).  We also categorize and index simplices in all dimensions for efficient traversal (Figure~\ref{fig:triangular}(d)).  In the rest of this section we formalize definitions (Section~\ref{sec:mesh:defs}), describe the subdivision of $(n+1)$-D prismatic meshes (Section~\ref{sec:mesh:prism}), and introduce the subdivision of $(n+1)$-D regular meshes as a special case of subdividing prismatic meshes (Section~\ref{sec:mesh:regular}).

\subsection{Definitions}
\label{sec:mesh:defs}

Formally, an \emph{$n$-simplex} is the convex hull of $n+1$ affinely independent points $a_0, a_1, \ldots, a_n$ in $\mathbb{R}^n$.  An \emph{$n$-simplicial complex} is the set of $k$-simplices ($k=0, 1, \ldots, n$); 
in this simplicial complex, any face of a simplex is part of the complex, and the intersection of any two simplices is either a lower-dimensional simplex or the empty set.

We define the \emph{$(n+1)$-D (simplicial) prism}\footnote{For ease of description, we limit the definition of prisms to those with simplicial bases.} as the extrusion of an $n$-simplex $a_0a_1\ldots a_n$ to one dimension higher.  Denoting the $\mathbb{R}^{n+1}$ coordinates of each point $a_i$ as $\mathbf{x}_i = (x_{i,0},x_{i,1},\ldots,x_{i,n})^\intercal$ with the identical last component $x_{i,n}$ for all $i$, the extruded prism includes another simplex $b_0b_1\ldots b_n$; the coordinates of each point $b_i$ are $(x_{i,0},x_{i,1},\ldots,x_{i,n}^\prime)^\intercal$, $x_{i,n} < x_{i,n}^\prime$.  Note that $a_i$ and $b_i$ share the same first $n$ coordinates and that the last coordinate is different.  In addition to the two simplicial bases, the prism includes $n$ edges $a_0b_0, a_1b_1, \ldots, a_nb_n$.  

We further define the \emph{$(n+1)$-D prismatic mesh} as the collection of $(n+1)$-D prisms obtained by extruding a simplicial mesh into one dimension higher.  Our goal is to tessellate the $(n+1)$-D prismatic mesh into a simplicial complex without adding new vertices.  

\subsection{Simplicial subdivision of (\emph{n}+1)-D prismatic meshes}
\label{sec:mesh:prism}

We first review the concept of \emph{staircase triangulation}~\cite{DeLoeraRS10} and then generalize the staircase triangulation to the simplicial subdivision of prismatic meshes.  Without loss of generality, we describe the case of $n=3$, the extrusion from an unstructured 2D triangular mesh to a 3D prismatic mesh, followed by its subdivision into a 3D tetrahedral mesh.  Assuming the input is given by a list of triangles (2-simplices), our algorithm extrudes each triangle into a prism in a new dimension; each triangular prism is further subdivided into three tetrahedra.  

\begin{figure}[h]
  \includegraphics[width=\linewidth]{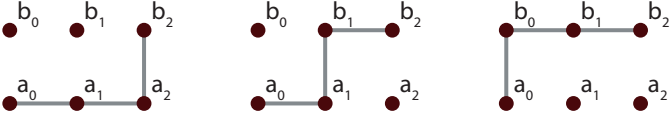}
  \caption{All monotone paths from $a_0$ to $b_2$ in the triangular prism $a_0a_1a_2-b_0b_1b_2$; each path corresponds to a staircase triangulation of the prism.}
  \label{fig:integral}
\end{figure}

\textbf{Staircase triangulation of a 3D triangular prism.}\quad As illustrated in Figure~\ref{fig:triangular}(d), a triangular prism may be subdivided into three tetrahedra.    Denoting the ``lower'' vertices as $a_0a_1a_2$ and the ``upper'' vertices as $b_0b_1b_2$, one may subdivide this triangular prism into three tetrahedra: $a_0a_1a_2b_2$, $a_0a_1b_1b_2$, and $a_0b_0b_1b_2$.  As documented by DeLoera et al.~\cite{DeLoeraRS10}, the vertex list of each tetrahedron corresponds to a \emph{monotone staircase} beginning with $a_0$ and ending with $b_2$ in Figure~\ref{fig:integral}; each vertex is immediately above or to the right of the previous vertex in the grid. 

\textbf{Staircase triangulation of an (\emph{n}+1)-D prism.}\quad The staircase triangulation can be generalized to higher dimensions, and an $(n+1)$-D prism may be subdivided into $n+1$ $(n+1)$-simplices without the introduction of new vertices.  First, we impose an ordering on the $n+1$ vertices in the lower and upper hyperplanes (and use the same ordering in both hyperplanes).  Second, we identify the $2(n+1)$ points of the prism with the grid $\{0, 1, 2, \ldots, n\}\times\{0, 1\}$.  Third, we compute all monotone paths\footnote{Monotone here means that both alphabets and subscripts are ascending.  For example, an edge like $b_0a_1$ or $a_1a_0$ cannot appear.} on the grid.  The staircase triangulation of 2D, 3D, and 4D prisms is listed in Table~\ref{tab:triangulation}.

\begin{table}[h]
  \caption{Monotone staircases and triangulation of 1D, 2D, 3D (triangular), 4D (tetrahedral), and 5D prisms.}
  \label{tab:triangulation}
  \centering
  \begin{tabular}{ccccc}
    1D          & 2D          & 3D          & 4D          & 5D \\
    \hline
    $a_0b_0$          & $a_0a_1b_1$       & $a_0a_1a_2b_2$    & $a_0b_0b_1b_2b_3$ & $a_0b_0b_1b_2b_3b_4$ \\
                      & $a_0b_0b_1$       & $a_0a_1b_1b_2$    & $a_0a_1b_1b_2b_3$ & $a_0a_1b_1b_2b_3b_4$ \\
                      &                   & $a_0b_0b_1b_2$    & $a_0a_1a_2b_2b_3$ & $a_0a_1a_2b_2b_3b_4$ \\
                      &                   &                   & $a_0a_1a_2a_3b_3$ & $a_0a_1a_2a_3b_3b_4$ \\
                      &                   &                   &                   & $a_0a_1a_2a_3a_4b_4$
  \end{tabular}\vspace{-0.15in}
  \includegraphics[width=\linewidth]{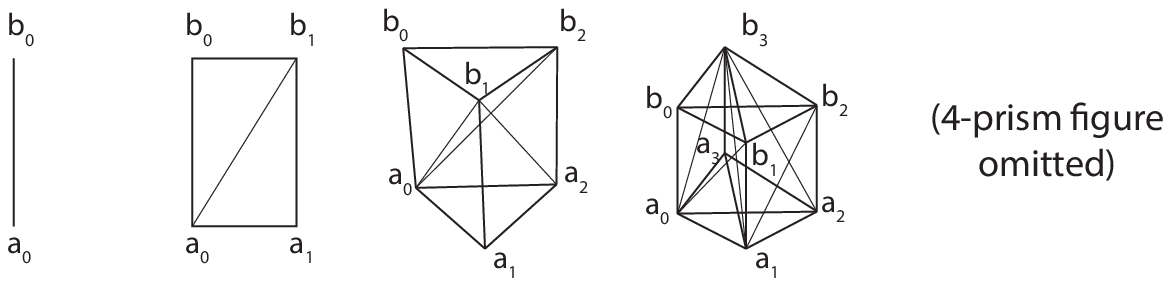}
\end{table}


\textbf{Staircase triangulation of (\emph{n}+1)-D prismatic mesh.}\quad The staircase subdivision method produces simplicial subdivisions along prism boundaries, given a global vertex ordering on a prismatic mesh.  In a 3D case in Figure~\ref{fig:triangular}(a-c), we assign a global order to each vertex and then subdivide each prism with staircase triangulation.  For example, the quadrilateral $a_1a_7b_7b_1$ is subdivided into two triangles $a_1a_7b_7$ and $a_1b_1b_7$ along the monotonous edge $a_1b_7$.

\textbf{Mesh element indexing.}\quad Each $k$-simplex in the subdivided $(n+1)$-D prismatic mesh can be one to one mapped to a tuple of integer ID, type, and timestep for traversing and compact storage.  Considering the extrusion along the new dimension for multiple layers of vertices (e.g., multiple timesteps), we use the same triangulation scheme for each layer and design an efficient indexing of simplices in all dimensions in the new mesh.  
For $k=3$, there are three types of 3-simplices: bottom, center, and upper tetrahedra (or type-I, type-II, and type-III tetrahedra), such that one can index each tetrahedron with a tuple of original triangle ID, type, and timestep.  The original triangle ID is the integer index of the triangle in the original mesh.  
For $k=2$, to uniquely index 2-simplices, we identify five unique types of faces: prism base, prism lower, prism higher, edge lower, and edge upper.  For example, the ``top'' triangle of a prism can be indexed by the ``bottom'' triangle of the same prism in the next timestep; triangles on quadrilaterals can also be indexed by neighboring prisms in the same layer.  As such, each 2-simplex can be uniquely indexed by the tuple of original triangle/edge ID, type, and timestep.  
Likewise, for $k=1$, we identify three unique types of edges; each can be indexed by the original vertex/edge ID, type, and timestep.  

\textbf{Mesh element queries.}\quad We provide functions such as  \texttt{vertices()}, \texttt{sides()}, and \texttt{side\_of()} that feature tracking algorithms that  can be used to query a mesh element in the extruded mesh.  The \texttt{vertices()} function returns the list of vertices of the given mesh element ID.  The \texttt{sides()} function provides a list of $(k-1)$-simplicial sides of the given \emph{k}-simplex, such as the triangular faces of a tetrahedron. The \texttt{side\_of()} function gives a list of $(k+1)$-simplices that contain the given $k$-simplex.  For example, a triangular face in 3D simplicial meshes is usually contained by two tetrahedra unless the face is on the boundary of the domain; likewise, a tetrahedron in a 4D simplicial mesh is usually shared by two pentachora (4-simplices).\footnote{A 4-simplex is equivalently referred as pentachoron, pentahedroid, pentatope, or tetrahedral pyramid in other literature; this paper uses the term pentachoron without loss of generality.}

\textbf{Ordinal and interval mesh elements.}\quad For ease of feature tracking, we further categorize $k$-simplices into \emph{ordinal} and \emph{interval} types.  A simplex is an ordinal type if each of its vertices resides in the same timestep in the extruded mesh; otherwise it is an interval type.  For example, the lower edge triangle (type-V face in Figure~\ref{fig:triangular}(d)) has two vertices in the lower layer and one vertex in the upper layer and is thus an interval type.  The type-I edge is ordinal because both vertices are in the same layer.  

There are two reasons to distinguish ordinal and interval types.  First, this distinction allows feature-tracking algorithms to consume data in a streaming and out-of-core manner for both in situ and post hoc processing, as discussed in the next paragraph.  Second, the distinction allows efficient slicing of output trajectories.  If one  needs only individual timesteps, rather than intervals, it is straightforward to select features identified in ordinal mesh elements. 

In a streaming pipeline, assuming data of each timestep $0, 1, \ldots, n_t-1$ is available in  ascending order, $n_t$ being the number of timesteps, we show that one can traverse all $k$-simplices while keeping a sliding window of two timesteps of data.  For each $i$th timestep, we first traverse ordinal types and then traverse interval types if $i<n_t-1$.  Because each interval type consists of vertices from adjacent timesteps,  both the $i$th and $(i+1)$th timesteps must be available in memory.  As a result, the discrimination of ordinal and interval types makes it possible to traverse every $k$-simplex without having all timesteps in memory simultaneously.

\textbf{Complexity.}\quad The space complexity of the subdivided $(n+1)$-prismatic mesh is of the order of mesh element count in the original $n$-simplicial mesh.
For example, in the case of $n=2$, we need to maintain lists of all triangles, edges, and vertices of the original triangular mesh.  We also maintain lists of sides and parents for all simplices in the original mesh, in order to accelerate the query of sides and parents in the extruded mesh.  The time complexity of querying a simplex and getting the vertex list of the simplex is constant.

\subsection{Simplicial subdivision of (\emph{n}+1)-D regular mesh}
\label{sec:mesh:regular}

\begin{figure}
  \includegraphics[width=\linewidth]{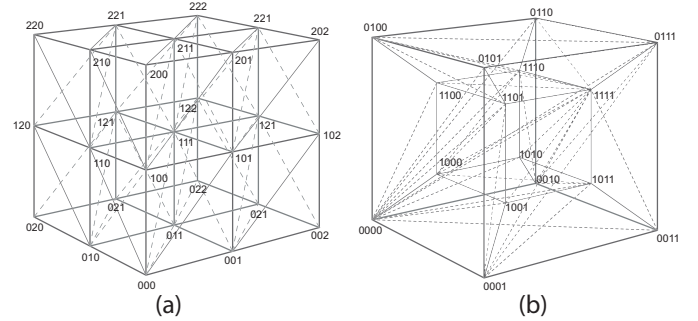}
  \caption{Regular simplicial meshes: (a) 3D mesh with eight cubes, each being subdivided to six tetrahedra, (b) 4D mesh with one single 4-cube, which is subdivided into 24 pentachoron.  Numbers encode both indices and coordinates of vertices.}
  \label{fig:regular}
\end{figure}

Subdividing a regular mesh is a special case of that in the preceding subsection but does not require maintaining a mesh data structure (e.g., lists of vertices and triangles).  We define the \emph{$(n+1)$-D regular simplicial mesh} as a simplicial subdivision of the $(n+1)$-D regular mesh without introducing additional vertices.  

\textbf{Recursive subdivision of (\emph{n}+1)-D regular mesh.}\quad One can recursively subdivide an $(n+1)$-D regular mesh based on the simplicial extrusion of an $n$-D regular simplicial mesh.  For $n=0$, the extruded mesh (1D regular grid) is already simplicial.  For $n\ge 1$, one can extrude cells in a $n$-D regular mesh into prisms and follow Table~\ref{tab:triangulation} to triangulate the prisms.  As a result, each $n$-D cube is subdivided in the same way into $n!$ congruent and disjoint $n$-simplices.  

\textbf{Precomputation of the subdivision for \emph{n}-D unit cube.}\quad In practice, we precompute the subdivision of the $n$-D unit cube for any $n$, which enables direct access to an $n$-D simplicial mesh without recursive subdivision.  For example, the unit 2-cube can be subdivided into two 2-simplices: 
\begin{equation}\nonumber
  \begin{array}{l}
    00, 01, 11\\
    00, 10, 11
  \end{array},
\end{equation}
where each 2-digit is the coordinate and ID of the vertex and each line is a 2-simplex.  By extruding the simplices, the simplicial subdivision of the unit 3-cube contains six tetrahedra: 
\begin{equation}\nonumber
  \begin{array}{l}
    000, 001, 011, 111\\
    000, 010, 011, 111\\
    \ldots\\
    000, 100, 110, 111
  \end{array},
\end{equation}
as is also illustrated in Figure~\ref{fig:regular}(a).  Likewise, the unit 4-cube can be subdivided into twenty-four ($4!=24$) pentachora, as illustrated in Figure~\ref{fig:regular}(b).
\begin{equation}\nonumber
  \begin{array}{ll}
    0000, 0001, 0011, 0111, 1111\\
    0000, 0010, 0011, 0111, 1111\\
    0000, 0001, 0101, 0111, 1111\\
    \ldots\\
    0000, 1000, 1100, 1110, 1111
  \end{array}.
\end{equation}
After the precomputation, each cube in the $n$-D regular mesh is subdivided in the same way.

\begin{figure}
  \includegraphics[width=\linewidth]{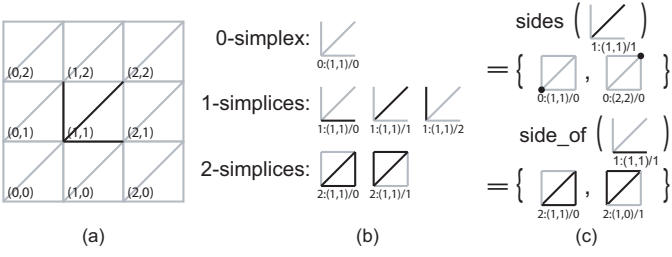}
  \caption{Indexing and querying mesh elements in regular simplicial mesh: (a) a 2D regular simplicial mesh; (b) indexing simplicial elements with $k:(i_0i_1)/\texttt{type}$, where $k$ is the dimensionality of the simplex, $(i_0i_1)$ is the corner coordinates of the cube that contains the simplex, and \texttt{type} is the unique simplex type ID; (c) \texttt{sides()} and \texttt{side\_of()} of a simplicial element.}
  \label{fig:tuple}
\end{figure}

\textbf{Mesh element indexing.}\quad We use the tuple of simplicial dimension $k$, corner coordinates, and the unique type ID to index a $k$-simplex in the $n$-D simplicial mesh.  The corner coordinates encode the location of the $n$-cube that contains the simplex.  The unique type ID is designed to encode a simplex within the cube; one cannot use other cubes to index the same simplex.  In a 2D case in Figure~\ref{fig:tuple}(a), although there are five edges (1-simplices) in each 2-cube, we have only three unique types of edges, because horizontal and vertical edges are always shared between neighboring cubes.  For example,  to index the top edge of cube (1, 1), we can use the cube (1, 2) to find the same edge.  Figure~\ref{fig:tuple}(b) enumerates all unique simplex types in 2-cubes.  

\textbf{Mesh element queries.} 
We also provide the \texttt{vertices()}, \texttt{sides()}, and \texttt{side\_of()} functions defined in the preceding subsection for $n$-D regular simplicial meshes.  The results of each function are precomputed for each unique type.  As illustrated in the 2D mesh in Figure~\ref{fig:tuple}(c), the sides of type-II 1-simplices (diagonal edges) include two vertices; two triangular cells contain the same type-I 1-simplices (horizontal edges).  

\textbf{Complexity.}\quad The space complexity of maintaining an $n$-D regular simplicial mesh is $O(n!)$, but $n$ does not exceed 4 for tracking features for 3D data.  Note that the space complexity does not grow with the  size of the regular grid, because precomputed unit-cube subdivisions are stored instead of an explicit list of mesh elements.  Such implicit mesh data structure allows queries of a simplex in constant time. 

%% file: critical.tex
\section{Tracking 0D features: critical points}
\label{sec:critical}

\begin{algorithm*}[!tb]
  \caption{Two-pass algorithm of tracking 0-, 1-, and 2- features---critical points, quantum vortices, and isosurfaces, respectively---with simplicial spacetime mesh.  $S$ is the set of simplices that test positive, \texttt{UF} being union-find.}\label{alg:tracking}
  \centering
  \begin{tabular}{c|c|c}
  \begin{minipage}{.3\linewidth}
    {\small
    \begin{algorithmic}
    \State $S \gets \varnothing$, \texttt{UF}$\gets \varnothing$
    \For{\textbf{each} \texttt{tri} $\in$ \texttt{simplices(2)}}
      \If{\texttt{test(tri)}} \Comment Eq.\ref{eq:inverse-linear}
        \State $S \gets S \cup$ \texttt{tri}
      \EndIf
    \EndFor
    \For{\textbf{each} \texttt{tet} $\in$ \texttt{simplices(3)}}
      \State $T \gets S \cap$ \texttt{tet.sides(2)}
      \State \texttt{UF.unite($T$)}
    \EndFor
    \end{algorithmic}
    }
  \end{minipage}
  &
  \begin{minipage}{.3\linewidth}
    {\small
    \begin{algorithmic}
    \State $S \gets \varnothing$, \texttt{UF}$\gets \varnothing$
    \For{\textbf{each} \texttt{tri} $\in$ \texttt{simplices(2)}}
      \If{\texttt{test(tri)}} \Comment Eq.\ref{eq:vortex}
        \State $S \gets S \cup$ \texttt{tri}
      \EndIf
    \EndFor
    \For{\textbf{each} \texttt{penta} $\in$ \texttt{simplices(5)}}
      \State $T \gets S \cap$ \texttt{penta.sides(2)}
      \State \texttt{UF.unite($T$)}
    \EndFor
    \end{algorithmic}
    }
  \end{minipage}
  &
  \begin{minipage}{.33\linewidth}
    {\small
    \begin{algorithmic}
    \State $S \gets \varnothing$, \texttt{UF}$\gets \varnothing$
    \For{\textbf{each} \texttt{edge} $\in$ \texttt{simplices(1)}}
      \If{\texttt{edge} intersects isosurface}
        \State $S \gets S \cup$ \texttt{edge}
      \EndIf
    \EndFor
    \For{\textbf{each} \texttt{penta} $\in$ \texttt{simplices(5)}}
      \State $T \gets S \cap$ \texttt{penta.sides(1)}
      \State \texttt{UF.unite($T$)}
    \EndFor
    \end{algorithmic}
    }
  \end{minipage}
  \end{tabular}
\end{algorithm*}

We describe here the use of simplicial meshes to track critical points in 2D and 3D vector fields.

\subsection{Assumptions and definitions}

We assume that the input $n$-dimensional time-varying vector field $\mathbf{v}: \mathbb{R}^{n+1} \to \mathbb{R}^n$ is piecewise linear.  The PL assumption implies that the vector field is defined on a simplicial spacetime mesh; each cell is an $(n+1)$-simplex.  For example, each cell in a 2D or 3D time-varying vector field is a tetrahedron (3-simplex) or pentachoron (4-simplex), respectively.  The $(n+1)$-simplicial spacetime mesh can be constructed based on an existing $n$-dimensional mesh, as detailed in Section~\ref{sec:hypermesh}.  Thus, 
$\mathbf{v}$ is $C^1$ continuous along any combination of spatial and temporal directions; that is, there exist $\mathbf{A}\in\mathbb{R}^{n\times(n+1)}$ and $\mathbf{b}\in\mathbb{R}^n$ for each $(n+1)$-simplex $S$ such that $\mathbf{v}=\mathbf{A}\mathbf{x}+\mathbf{b}$, $\mathbf{x}\in S$.

We further assume that the time-varying vector field is \emph{generic}.  This means that vector values on each vertex $i$ are nonzero ($\mathbf{v}_i\neq\mathbf{0}$) and that vectors at verticies of any $k$-simplex ($k=0, 1, \ldots, n+1$) are affinely independent.  Thus, critical points in generic vector fields may be found in the interior of $n$-simplices instead of on cell boundaries.  In the end of this subsection we discuss the relaxation of the generic assumption by using the simulation of simplicity~\cite{EdelsbrunnerM94a} technique.  

A \emph{(spacetime) critical point} $\mathbf{x}_c\in\mathbb{R}^{n+1}$ is the location where the vector value $\mathbf{v}(\mathbf{x}_c)$ is zero.  We  focus on critical points that are nondegenerate; meaning that the (spatial) Jacobian $\mathbf{J}_{\mathbf{v}}$ at $\mathbf{x}_c$ is nondegenerate.  Based on the eigensystem of $\mathbf{J}_{\mathbf{v}}$, the critical point $\mathbf{x}_c$ can be further categorized into various types such as sources, sinks, and saddles.  In the case that $\mathbf{v}$ is the gradient field of a scalar field, the critical point types are maxima, minima, and saddles. 

A \emph{critical point trajectory} (or simply trajectory) is a locus of critical points in space and time, which are 1D curves embedded in $\mathbb{R}^{n+1}$.  Because $\mathbf{v}$ is PL, critical point trajectories are PL parametric curves, and the intersection with each cell is a line segment.  Critical point trajectories can be \emph{sliced} into a set of critical points at an arbitrary time $t_0$ by intersecting the hyperplane $t=t_0$.  In the following sections we discuss methods for reconstructing critical point trajectories.

\subsection{Two-pass critical point trajectory reconstruction}

\begin{figure}
  \includegraphics[width=\linewidth]{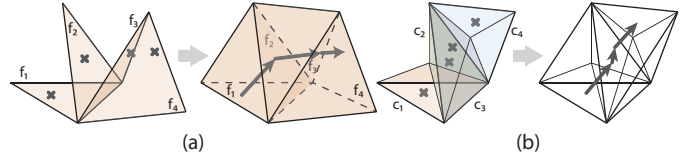}
  \caption{Two-pass critical point trajectory reconstruction for 2D and 3D vector fields with 3D (a) and 4D (b) spacetime simplicial meshes, respectively.  The first pass tests whether triangular/tetrahedral sides intersects the trajectory, and then the second pass associates every pair of intersected sides if they share the same tetrahedron/pentachoron.}
  \label{fig:tracking}
\end{figure}

As illustrated in Figure~\ref{fig:tracking}, we use a two-pass algorithm to reconstruct critical point trajectories from $\mathbf{v}$.  Without loss of generality, we describe this algorithm with 2D time-varying vector fields ($\mathbf{v}: \mathbb{R}^3\to\mathbb{R}^2$).  In the first pass, we iterate each triangular face (2-simplex) to determine whether a trajectory intersects the face by solving the inverse linear interpolation problem: 
\begin{equation}\label{eq:inverse-linear}
  \left(\begin{matrix}u_0 & u_1 & u_2\\v_0 & v_1 & v_2\\1 & 1 & 1\end{matrix}\right)
  \left(\begin{matrix}\mu_0\\\mu_1\\\mu_2\end{matrix}\right) = 
  \left(\begin{matrix}0\\0\\1\end{matrix}\right),
\end{equation}
where $(\mu_0, \mu_1, \mu_2)^\intercal$ are the normalized barycentric coordinates of the trajectory intersection with the face and $(u_0, v_0)^\intercal$, $(u_1, v_1)^\intercal$, and $(u_2, v_2)^\intercal$ are the vector values on the three vertices of the 2-simplex.  If $\mu_j \in [0,1]$ for all $j\in\{0, 1, 2\}$, then the triangular face is punctured by a trajectory, and the spacetime coordinates of the critical point can be calculated.
In the second pass, we iterate over each tetrahedral cell (3-simplex) to associate its sides that are punctured by trajectories, because one can prove that each tetrahedron intersects up to one trajectory.  Complete trajectories can be constructed by pairing every two punctured triangular faces of the same tetrahedron.  

In general, for arbitrary dimensionality $n$, the two-pass algorithm to reconstruct critical point trajectories in $\mathbf{v}: \mathbb{R}^{n+1} \to \mathbb{R}^n$ is the following.  The first pass iterates over each $n$-simplex to determine whether the simplex intersects a trajectory based on Equation~\eqref{eq:inverse-linear}.  The second pass iterates each $(n+1)$-simplex and pairs its sides ($n$-simplices) that intersect a trajectory.  The two-pass algorithm can be easily parallelized with both distributed and GPU parallelism, as discussed in the following sections.  

\begin{figure}
  \includegraphics[width=\linewidth]{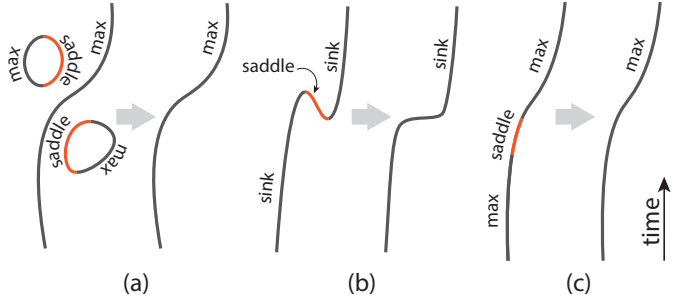}
  \caption{Example of (a) filtering, (b) simplification, and (c) smoothing of critical point trajectories.  Colors indicate different critical point types.}
  \label{fig:simplification}
\end{figure}

The output reconstructed critical point trajectories are closed curves in spacetime; they either end on domain boundaries or are loops.  Within each curve, the critical point type may alternate, and there may be multiple monotone intervals with respect to time.  
For example, as illustrated in Figure~\ref{fig:simplification}(a), each loop characterizes a maximum-saddle pair in the gradient field; the maximum-saddle pair establishes and annihilates simultaneously.  In Figure~\ref{fig:simplification}(b), we see the birth of a saddle-sink pair; the saddle further merges with another sink soon after the birth.  One can further simplify and filter trajectories based on their attributes, as discussed in the following sections.

\subsection{Robustness}
\label{sec:robust-cp}

The  two-pass algorithm assumes that PL vector fields are generic, an assumption that often does not hold for real-world data.  For example, gradients of an integer-valued image may be exactly zero at vertices; gradients based on central-differences, which are rational numbers, may be affinely dependent, causing nongeneric situations.  In fluid flows, nonslip conditions lead to zero velocities on boundaries.  
Ideally, there would be a guarantee that each $(n+1)$-simplex has at most one pair of intersected sides, but this is not true for nongeneric cases.
A critical point may reside on the boundary of the cell, causing numerical instabilities during the test of $\mu_k \in [0,1]$ in Equation~\eqref{eq:inverse-linear}.  As a result, the critical point may or may not be detected in the current $n$-simplex; the same critical point may be detected by neighboring cells, causing nonrobust and noncombinatorial tracking results, as illustrated in Figure~\ref{fig:robust}.  

\begin{figure}
  \includegraphics[width=\linewidth]{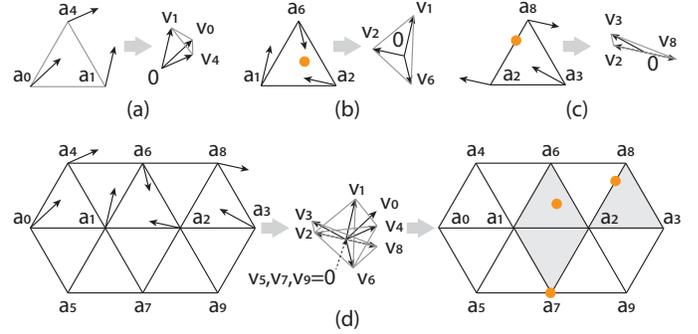}
  \caption{Example of robust critical point test in a 2D mesh; a triangle $a_ia_ja_k$ tests positive if the origin point $\mathbf{0}$ is in the interior of the simplex $\mathbf{v}_i\mathbf{v}_j\mathbf{v}_k$, which consists of vector values at vertices.  The point-in-simplex test is robust and combinatorial based on SoS.}
  \label{fig:sos-critical}
\end{figure}

We use Simulation of Simplicity~\cite{EdelsbrunnerM94a} (SoS) to compute critical point trajectories robustly and combinatorially in nongeneric vector fields.  As proved by Bhatia et al.~\cite{BhatiaGWBP14} with the Brouwer degree theory, a critical point exists in the $n$-simplex $\{\mathbf{x}_0, \mathbf{x}_1, \ldots, \mathbf{x}_n\}$ if and only if $\mathbf{0}$ lies in the interior of the convex hull of $\{\mathbf{v}_0, \mathbf{v}_1, \ldots, \mathbf{v}_n\}$, where $\mathbf{v}_j$ ($j=0, 1, \ldots, n$) is the vector value on each vertex $\mathbf{x}_i$ (illustrated in Figure~\ref{fig:sos-critical}(a-c)).  As a result, the critical point test is reduced to the point-in-simplex predicate, which can be determined by the sign of the determinant of the matrix $(\mathbf{v}_0, \mathbf{v}_1, \ldots, \mathbf{v}_n, \mathbf{1})$.  

In nongeneric cases, such as a critical point on the boundary of the simplex, the SoS prevents the determinant from becoming zero by adding a symbolic perturbation to each component of the matrix.  The perturbation, namely, $\epsilon$-expansion, is a function of an arbitrarily small number $\epsilon$; the form of the perturbation is determined by the vertex and dimension ordering.  As a result, if a critical point lies on the boundary that is shared by two or more simplices, SoS implicitly enforces a consistent choice to exclusively associate the critical point with  one of the simplices, as illustrated in Figure~\ref{fig:sos-critical}(d).

\subsection{Critical point trajectory filtering, simplification, and smoothing}

We provide three postprocessing approaches to help users filter, smooth, and simplify trajectories that result from tracking critical points.  

\textbf{Filtering.}\quad One can filter results based on the trajecxtory attributes---time duration, topology, persistence, scalar value, if applicable.  Figure~\ref{fig:simplification}(a) is an example of filtering loops in the gradient of a scalar field.  Typically, a loop exists when a small transient bump appears, introducing a saddle-extremum pair.  Such a loop may be filtered out based on the  time duration or persistence of the loop.  Likewise, trajectories can be filtered based on other attributes, as demonstrated in the following sections.

\textbf{Simplification.}\quad One can also simplify trajectories that change directions frequently in time based on a threshold of persistence in time.  As illustrated in Figure~\ref{fig:simplification}(b), for example, a saddle-sink pair is born right before the saddle merges with another sink.  Because the saddle may be caused by noise, we provide the simplification function to eliminate the saddle and merge the trajectory as a consistent sink type.  

\textbf{Smoothing.}\quad Although our trajectory reconstruction algorithm is robust, the evaluation of Jacobians $\mathbf{J}_{\mathbf{v}}$ may subject to numerical instabilities, causing inconsistent critical point types along trajectories.  
One can smooth the types based on a window-shifting approach.  We iterate each point in the trajectory and check whether the current critical point type is consistent with both the precedent and antecedent.  If an inconsistency is identified, we mark the inconsistency location and modify the type after the iterations, as illustrated in Figure~\ref{fig:simplification}(c).  The window size depends on the application and analysis needs.  For example, we set the half-window size to be two consecutive timesteps in our experiments.

\subsection{Evaluation and verification with synthetic data}
\label{sec:critical:synth}

We validate the effectiveness and evaluate the robustness of our critical point tracking approach with synthetic data.  

\begin{figure}
  \includegraphics[width=\linewidth]{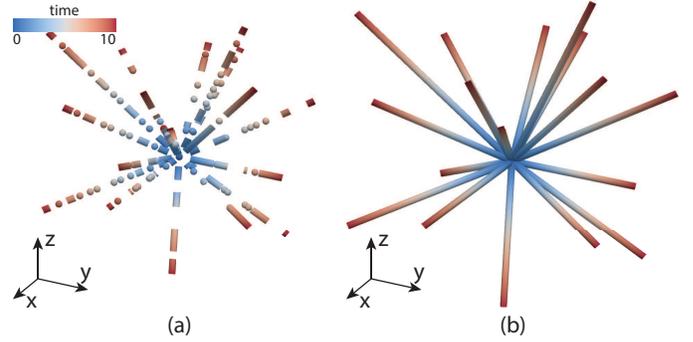}
  \caption{Nonrobust (a) versus robust (b) tracking of critical points in 20 instances of the 3D moving minimum synthetic data, each with the same origin but random rational directions.  Color encodes time in the figure.}
  \label{fig:nonrobust}
\end{figure}

\textbf{3D moving minimum.}\quad We use the following scalar function to synthesize 3D time-varying scalar field data with the known position of the single minimum to evaluate the numeric robustnesss of our method:  
\begin{equation}
  f(\mathbf{x}, t) = \Vert \mathbf{x} - (\mathbf{x}_0 + \mathbf{d}\cdot t) \Vert^2,
\end{equation}
where $(\mathbf{x}, t)\in \mathbb{R}^{n+1}$ are the spatiotemporal coordinates and $\mathbf{x}_0$ and $\mathbf{d}$ are arbitrary vectors in $\mathbb{R}^n$.  As a result, the trajectory of the single minimum $\mathbf{x}_c(t)$ in the data is 
\begin{equation}
  \mathbf{x}_c(t) = \mathbf{x}_0 + \mathbf{d}\cdot t.
\end{equation}
In Figure~\ref{fig:nonrobust}, we synthesized 20 different instances with the same $\mathbf{x}_0=(10, 10, 10)^\intercal$ but with different $\mathbf{d}$.  The scalar function $\mathbf{f}$ is discretized into a $21\times21\times21$ grid, which represents a $(0, 0, 0)\times(20, 20, 20)$ domain.  Each component of the moving direction $\mathbf{d}$ is a random rational number such that the trajectory must intersect at least one grid point including $\mathbf{x}_0$, causing degenerate cases similar to that of Figure~\ref{fig:robust}.  In Figure~\ref{fig:nonrobust}(a), because the grid point of $\mathbf{x}_0$ is shared by multiple pentachora in the spacetime mesh, the number of tetrahedra that numerically test positive for containing a critical point ranges from zero to tens.  Because the degenerate cases cause ambiguity in tracing, trajectories in the figure appear dashed and isolated.  In Figure~\ref{fig:nonrobust}(b), our robust detection approach guarantees that each critical point is exclusively associated with a tetrahedron such that trajectories can be tracked robustly without any ambiguity.

\begin{wrapfigure}{r}{3.2cm}
  \includegraphics[width=\linewidth]{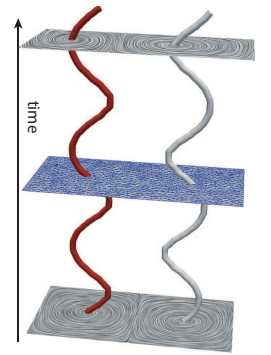}
  \caption{Critical point tracking in 2D unstructured mesh double gyre flow.  Color encodes ID of trajectories.}
  \label{fig:double-gyre}
\end{wrapfigure}
\textbf{2D double gyre flow.}\quad We demonstrate critical point tracking in a 2D unstructured mesh with the double gyre function, which is widely used to study Lagrangian coherent structures~\cite{double-gyre}.  
The double gyre function is defined in the domain of $[0,0]\times[2,1]$, and we generate a triangular mesh with 2,098 triangles and 1,100 vertices to demonstrate the results.  Figure~\ref{fig:double-gyre} visualizes critical trajectories in $t\in[0, 40]$; the timespan $\Delta t$ between adjacent timesteps is 0.1.  The color of each trajectory is categorical and encodes the unique ID of the trajectory.  Slices at $t=0$ and $t=2.8$ visualize flow directions with line integral convolution (LIC), and the slice at $t=1.5$ visualizes the 2D mesh.  As shown in the figure, two critical points move back and forth along the $x$-axis periodically inside the 2D domain.  Because the double gyre function is analytical, we verified that the vector field is exactly zero at all points in the trajectories, and all critical points are identified by our method.  

\begin{figure}
  \includegraphics[width=\linewidth]{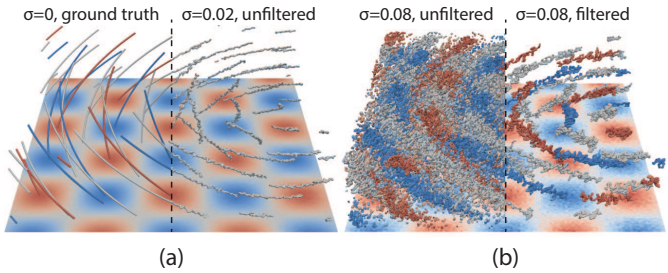}
  \caption{Critical point tracking and filtering in the 2D synthetic spiral woven data. (a) left: no noise injection, right: $\sigma$ = 0.02 without filtering, (b) left: $\sigma$ = 0.08 without filtering, right: $\sigma$ = 0.08 with filtering.}
  \label{fig:woven}
\end{figure}
\textbf{2D spiral woven with perturbations.}\quad We design the spiral woven function $f$ to evaluate our critical point tracking method with the presence of noise: 
\begin{equation}
  f(x, y) = \cos(x\cos t - y\sin t) \sin(x\sin t + y\cos t).
\end{equation}
This function has a finite number of critical points including minima, maxima, and saddles for a bounded 2D domain.  The critical points rotate around the origin at a fixed angular speed over time.  We discretized the function into a $128^2$ grid and injected Gaussian noise with two different standard deviations into the data, as shown in Figure~\ref{fig:woven}.  Note that the range of the data is $[-1, 1]$, and thus perturbation of $\sigma=0.02$ and $\sigma=0.08$ introduces up to 3\% and 12\% relative error, respectively, to the data in the 3-sigma limits.  Because the noise injection produces many artificial bumps in the data, the output critical point trajectories contain artifacts.  Artifacts such as small loops can be removed through trajectory filtering, as discussed in the preceding subsection. 

\subsection{Case studies with applications}

\begin{figure*}
  \includegraphics[width=\linewidth]{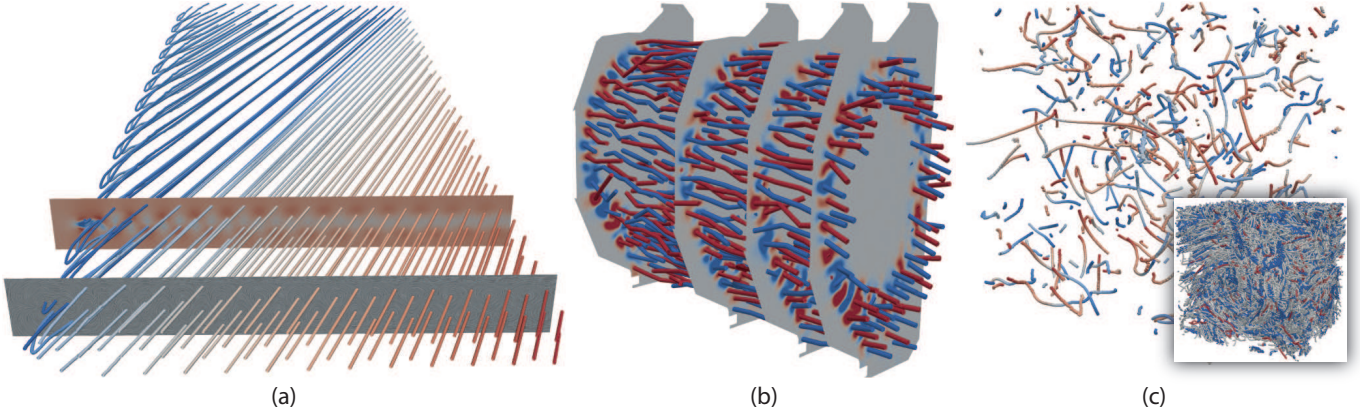}
  \caption{Case studies of critical point tracking. 
  (a) 2D flow-past-cylinder vector field (data courtesy of Weinkauf and Theisel~\cite{weinkauf10c}); color encodes the ID of each trajectory. The two 2D slices at \emph{t} = 60 and \emph{t} = 200 each visualize the velocity direction and magnitude with line integral convolution image and pseudo colors, respectively.
  (b) XGC fusion simulation (data courtesy of Dominski et al.); color of trajectories encodes whether a critical point is a maximum (red) or minimum (blue).
  (c) Turbulent vortices (data courtesy of Ma); color encodes time (0 $\leq$ \emph{t} $\leq$ 30) of all maximal trajectories in the large image; color encodes critical point types (blue for minima, white for saddles, and red for maxima) in the small image.}
  \label{fig:cases-critical}
\end{figure*}

We demonstrate use cases of our critical point tracking methods in science applications. 

\textbf{Fluid dynamics.}\quad We track vector field critical points to visualize the vortex streets in a 2D flow-past-cylinder simulation (Figure~\ref{fig:cases-critical}(a)).  The data are available on a uniform grid with the resolution of $400\times50$ for 1,001 timesteps.  We use the average velocity as the frame of reference and track all critical over time.  Based on the trajectories, we can see that critical points are created in pairs past the cylinder and move evenly toward the other side of the domain.

\textbf{Tokamak simulations.}\quad We use trajectories of scalar field critical points to characterize the dynamics of blobs in tokamak fusion plasma simulations (Figure~\ref{fig:cases-critical}(b)).  Understanding blobs---regions of high turbulence that could damage the tokamak---is critical for reactor design and future power generation.  Assuming each blob corresponds to a maximum/minimum in the preconditioned and normalized electron density field, we reconstruct trajectories of critical points in a single poloidal plane consisting of 56,980 vertices and 112,655 triangles in an XGC particle-in-cell simulation~\cite{xgc}.  These trajectories can enable further analyses of the physical properties of blobs.

\textbf{Turbulent vortices.}\quad We track and visualize scalar field critical points in a publicly available turbulent vortex dataset~\cite{tvdr}, which contains 100 timesteps of $128^3$ scalar value data characterizing vorticity magnitudes.  We visualize results for the first 30 timesteps (Figure~\ref{fig:cases-critical}(c)) to avoid overplotting.  There are 549 maximal trajectories out of 85,911 trajectories in all types (minima, saddles, and maxima).  In the visualization, one can see how high-turbulent locations move over space and time.  


\subsection{Limitations}

First, in the case of critical point tracking in scalar fields, one has to use the derived gradient fields as the input.  The spacetime PL gradient field implies $C^2$ continuity of the original scalar field; this is different from topology methods, which assume a PL scalar field and identify critical points at vertices of the mesh~\cite{Ban67}.  The outputs may be distorted because of the smoothness of differentiation kernels used for gradient derivation.  That said, the output trajectories are as accurate as the quality of the gradient field and how close the scalar field is to $C^2$ continuity.  In addition, the fidelity of critical point trajectories is related to the spacetime resolution of the inputs.  Downsampling the data may lead to an oversimplified topology of the trajectories.  

Second, the determination of critical point types is subject to numerical instabilities.  Because Jacobians are usually estimated numerically, in extreme cases  the signs of eigenvalues may change when perturbation is introduced.  We therefore introduced a smooth filter to heuristically correct classification errors, but one can also remediate the problem based on domain-specific knowledge.

\begin{wrapfigure}{r}{2.5cm}
  \includegraphics[width=\linewidth]{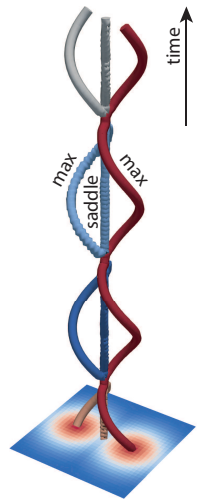}
  \caption{Concurrent merges and splits of three critical points.  Color encodes trajectory ID.}
  \label{fig:merger}
\end{wrapfigure}
Third, the PL assumption prevents the native identification of higher-order critical points~\cite{TricocheSH00, WeinkaufTSHS05}.  An example of higher-order cases is illustrated in Figure~\ref{fig:merger}, where two local maxima periodically merge and split.  However, the result includes a persisting trajectory that characterizes one of the maxima and a number of maxima-saddle loops periodically.  At the time of merging, ideally three critical points (two maxima and one saddle) should merge and then split into another three critical points.  Because at most one trajectory is assumed to intersect each cell, our method cannot identify the ``3-in-3-out'' event over time.  We leave the study of higher-order dynamics of critical points to future work as well.

\subsection{Comparison with existing critical point tracking algorithms}

Our approach offers three improvements over existing approaches that use spacetime meshing~\cite{TricocheWSH02, GarthTS04}.  First, our method uses simplicial meshes instead of prismatic meshes in 4D spacetime.  Because each of our simplicial cells intersects at most one singularity, our method consistently avoids ambiguity when multiple trajectories intersect a prism. Second, the PL assumption makes it easier to localize zero crossings in simplicial cells than in prismatic cells, which may result in multiple pairs of intersections on the cell boundary and cause challenges in the parity test.  Third, our simplicial mesh enables the robust tracking of critical points based on SoS, producing combinatorial and consistent results when nongeneric cases occur.

Compared with approaches based on feature flow fields~\cite{TheiselS03, WeinkaufTGP11, KleinE07, ReininghausKWH12}, our method is numerically robust and computationally scalable.  FFFs are vector fields that characterize the movement of critical points such that the trajectories can be computed as tangent curves in FFFs.  First, in comparison with FFF, which requires numerical approximations of the gradients of vector fields, our method does not require such transformation.  Numerical integration in FFF, such as Runge--Kutta methods, further introduces error in curve tracing.   
Second, our two-pass algorithm can be embarrassingly parallelized, while computing streamlines in distributed and parallel environments in general can be difficult to scale~\cite{PugmirePG12}.

%% file: vortex.tex
\section{Tracking 1D features: quantum vortices}
\label{sec:vortex}

We demonstrate the use of simplicial spacetime meshing to track 1D topological defects (quantum vortices) in 3D complex-valued scalar fields produced by superconductivity, superfluidity, and Bose--Einstein condensate simulation data.  Unlike critical points in vector fields, vortices are 1D curves in individual timesteps, and the trajectories of vortices are 2D surfaces embedded in 4D spacetime.

\subsection{Two-pass vortex curve and surface reconstruction}

Vortex tracking can be achieved by a two-pass reconstruction algorithm, as outlined in Algorithm~\ref{alg:tracking} (middle).  We review here how vortices intersect 2- and 3-simplices, and we then introduce the two-pass vortex surface reconstruction.  

\begin{figure}
  \includegraphics[width=\linewidth]{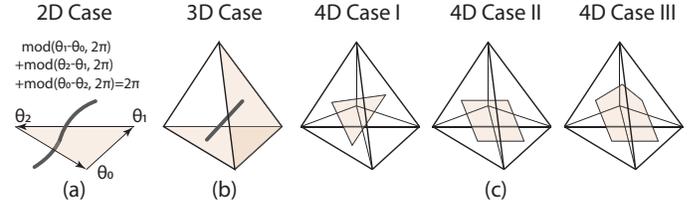}
  \caption{Cases of a quantum vortex intersecting a 2-, 3-, and 4-simplices, respectively.}
  \label{fig:vortex1}
\end{figure}

\textbf{Contour integral test on triangular faces.}\quad We test whether each spacetime 2-simplex intersects a vortex based on the definition in Equation~\eqref{eq:vortex}.  As shown in Figure~\ref{fig:vortex1}(a), a triangular face intersects a vortex at most once.  The contour integral can be calculated by accumulating phase shifts along each edge: 
\begin{equation}
  \oint_C \nabla \theta(\mathbf{x}) \cdot d\mathbf{l} = \sum_{i=0}^2 \Delta \theta_{i,j}, i=0, 1, 2, j=i+1\  \textrm{mod}\ 3,
\end{equation}
where the phase shift $\Delta \theta_{i,j}$ is the modulo of $\theta_j-\theta_i$ and $2\pi$; $\theta_i$ and $\theta_j$, respectively, are the phase angle at the $i$th and $j$th vertex of the triangular face.  The test is positive if the contour integral equals $\pm 2\pi$.  Because the modulo is less than $2\pi$, we assume that the data resolution in both spatial and temporal dimensions is sufficiently fine  that the phase difference between each pair of adjacent vertices is less than $2\pi$.  Scientists would need to refine the spatiotemporal resolution of simulations if the discretization is too coarse.

\textbf{Vortex curve reconstruction.}\quad One can reconstruct the 1D topology of vortices in the $\mathbb{R}^3$ subspace of individual timesteps.  Because vortices are closed curves In $\mathbb{R}^3$, each closed volume has an even number of intersections on boundaries; each tetrahedral cell has up to one pair of intersected faces~\cite{GuoPG17}, as illustrated in Figure~\ref{fig:vortex1}(b).  As a result, vortex curves can be reconstructed by associating intersected faces that share the same tetrahedral cells.  The process involves  two passes:  first iterating all triangular faces and then scanning tetrahedral cells that have intersected faces. 

\textbf{Vortex surface reconstruction.}\quad The two-pass curve reconstruction can be generalized to 4D spacetime to characterize the trajectory of vortex curves as surfaces.  In 4D, each pentachoron
has five tetrahedral faces; each tetrahedron may intersect a vortex.  Since tetrahedron shares triangular sides in the pentachoron, the number of triangular sides that test positive may be 3, 4, or 5 (corresponding to Case I, II, or III in Figure~\ref{fig:vortex1}).  Each intersection sits on the same 2-manifold of the vortex surface.  In the reconstruction of the vortex surface, the first pass tests all triangular faces in 4D spacetime, and then the second pass joins all triangular faces that test positive and share the same pentachoron.  



\textbf{Benefits.}\quad The use of spacetime simplicial meshing simplifies and improves on previous implementations based on cubic meshes~\cite{GuoPPKG16} and prismatic meshes~\cite{GuoPG17}.  In the former study, ambiguity exists when two vortices penetrate the same cube.  In the latter study,  in order to eliminate ambiguities with cubic cells, each spatial cube is tessellated into six tetrahedra such that each tetrahedron intersects up to one vortex line.  Although mesh elements in individual timesteps are simplices, prisms are used in spacetime, causing two problems: (1) two different functions are needed to test triangular and quadrilateral faces of a prism, and (2) ambiguity is still possible with prismatic cells.  With simplicial spacetime meshes, we need only one  function to test triangular faces, and there is no ambiguity.

\subsection{Case study of a superconductivity simulation}

\begin{figure*}
  \includegraphics[width=\linewidth]{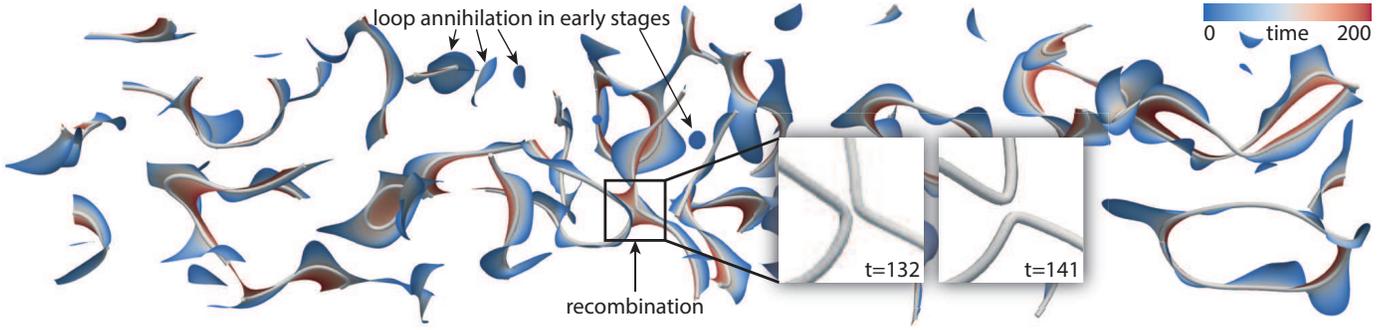}
  \caption{Trajectory surfaces of quantum vortices in a 3D superconductivity simulation data with 200 timesteps; color encodes the timestep.  Vortices at timestep 132 are visualized with tubes.  The two zoomed subfigures visualize how a pair of vortices exchanges parts before and after the recombination event.}
  \label{fig:surface}
\end{figure*}

Figure~\ref{fig:surface} demonstrates vortex tracking results\footnote{We have cross-validated the results by comparing with our previous publications.  Specific to this simulation, a local gauge transformation is used for numerical treatment.  See more details in~\cite{GuoPPKG16,GuoPG17}.} of a time-dependent Ginzburg--Landau superconductivity simulation, produced by the finite-difference partial differential equation solver GLGPU~\cite{SadovskyyKPKG15}.  In this case, the resolution of the mesh is $512\times128\times64$, and the number of the timesteps is 200.  As a result, 6.5M out of 251M spacetime triangular faces tested positive for encircling a vortex, forming a set of disjoint surfaces.  

From the scientific perspective, the dynamics of vortices determine all electromagnetic properties of the superconducting materials, and recombinations of vortices are directly related to energy dissipation~\cite{GlatzVKC16}.  The surface-based visualization produced by our tools enables scientists to investigate the time-varying features with a single image.  The reconstructed surface also makes it possible to derive the moving speed of each vortex when there are no topological changes; the moving speed has positive correlation with the voltage drop, which is critical to the material design. 


%% file: contour.tex
\section{Tracking 2D features: isosurfaces}
\label{sec:contour}

Isosurface tracking in FTK is also based on simplicial spacetime meshes.  As demonstrated by Bhaniramka et al.~\cite{BhaniramkaWC04} and Ji et al.~\cite{JiSW03},  isosurfaces can be tracked in 3D scalar fields by extracting and slicing levelsets in $\mathbb{R}^4$ regular meshes.  While the major complexities of previous efforts---disambiguation of isosurfaces intersecting the same hypercube---may be resolved by triangulation, our implementation with simplicial spacetime meshing intrinsically avoids ambiguities in a consistent manner and scales to larger computing resources with the FTK framework.  

\subsection{Two-pass isovolume reconstruction in spacetime scalar fields}

We formalize the isosurface tracking problem as the reconstruction of isovolumes in the time-varying scalar field $f: \mathbb{R}^{n+1}\to\mathbb{R}$, isovolumes being the solution of $f=c$, where $c$ is the isovalue.  In general, the isovolume is an $(n-1)$-dimensional object embedded in $\mathbb{R}^{n+1}$.  With $n=3$, the object can be further sliced into 2D surfaces with fixed time values.  We assume that $f$ is generic; that is, the scalar value $f_i\neq c$ for each vertex $i$ and scalars at vertices of any $k$-simplex ($k=1, 2, \ldots, n+1$) are affinely independent.  Thus, each edge (2-simplex) in the spacetime mesh intersects the level set at most once; the edge neither resides on the level set nor intersects the it at vertices.  Similar to the robust critical point tracking method in Section~\ref{sec:critical}, we use symbolic perturbations~\cite{EdelsbrunnerM94a} to relax the generic assumption and to reconstruct 4D level sets in a robust and consistent manner for real application data.  

As shown in Algorithm~\ref{alg:tracking}, the reconstruction consists of two passes for isovolume reconstruction in 4D: the edge pass and pentachoron pass.  In the first pass, we check whether every edge (2-simplex) intersects a level set.  In the second pass, we iterate every pentachoron; edges that intersect the level set are associated and labeled with the same identifier.  

\begin{figure}[h]
  \includegraphics[width=\linewidth]{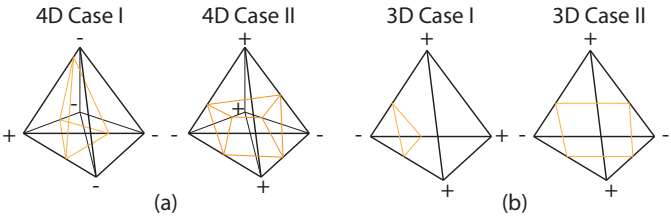}
  \caption{Cases of an isovolume intersecting a pentachoron (a) and a tetrahedron (b).  Signs indicate whether the scalar value on a vertex is greater or less than the given isovalue.}
  \label{fig:contour}
\end{figure}

In the case of a 3D time-varying scalar field, the output isovolumes are 3D objects and can be represented as a tetrahedral grid, which can be further sliced into isosurfaces in individual timesteps for visualization and analysis.  As illustrated in Figure~\ref{fig:contour}(a), the intersection between a pentachoron and an isovolume has only two distinct cases.  We use the positive and negative sign, respectively, to represent whether the scalar value on a vertex is greater or less than the threshold.  In the 4D case I ($+----$ or other permutations), one of the vertices has a different sign than other vertices do; in this case, the isovolume inside the pentachoron is the single tetrahedron consisting of the four intersections.  In the 4D case II ($++---$ or other permutations), the pentachoron has six intersections, which form a polytope and can be triangulated into three tetrahedra, as explained below. 

Without loss of generality, we show that the second case $++---$ leads to an 3-polytope that can be tessellated into three tetrahedra.  In the five tetrahedral sides of the pentachoron, we find three tetrahedra with $++--$ and the other two tetrahedra with $+---$.  As illustrated in Figure~\ref{fig:contour}(b), in the 3D case I ($+---$), the intersection is a triangle; in the 3D case II ($++--$), the intersection of the tetrahedron and the isovolume leads to a coplanar quadrilateral.  As a result, we have two triangles and three quadrilaterals, forming a prism-like polytope that resides in the same 3D subspace.  With the same staircase triangulation method described in Section~\ref{sec:mesh:prism}, we can decompose the polytope into three non-overlapping tetrahedra in a combinatorial manner.

\subsection{Robustness}

The robustness of this method is guaranteed by the assumption that the function is generic; that is, the scalar value of each vertex is either greater or less than the isovalue.  For real problems that usually do not observe this assumption, degenerate cases may appear.  For example, should the scalar value of a vertex exactly equal the isovalue, the intersection may or may not be identified by other edges that share the same vertex, leading to unstable results.  Should an edge reside on an isosurface, every point on the edge is part of the isosurface, leading to numerical instabilities.  

In nongeneric cases, we ensure the robustness with the same SoS programming technique~\cite{EdelsbrunnerM94a} used for robust critical point tracking (described in Section~\ref{sec:robust-cp}), and we regard the isosurface/edge intersection test as a special case of critical point detection in a 1D vector field.  Thus, if an intersection is on any vertex of the edge, the SoS prevents the divisor from becoming zero by adding a symbolic perturbation, such that the intersection is exclusively associated with one of the edges that share the same vertex.

\subsection{Case studies}

\begin{figure}
  \includegraphics[width=\linewidth]{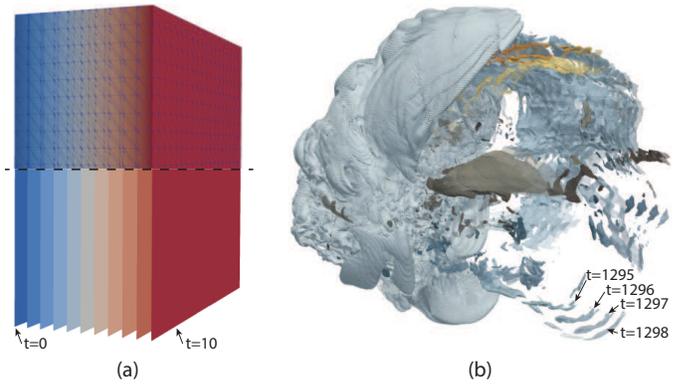}
  \caption{Tracking isosurfaces: (a) projected isovolume (top) and isosurfaces (bottom) of synthetic data and (b) isosurfaces of supernova data.  Colors in (a) and (b) encode time and surface ID, respectively.}
  \label{fig:isovolume}
\end{figure}

We demonstrate our isovolume reconstruction with both synthetic and simulation data in Figure~\ref{fig:isovolume} and below.

\textbf{Synthetic data.}\quad We show in Figure~\ref{fig:isovolume}(a) the reconstruction of the isovalue at $f=0$ for the function $f(x, y, z, t) = x - \alpha t$ on a $21\times21\times21$ grid and 12 timesteps, with $\alpha=0.9$.   Since the closed form of the surface trajectory ($x=\alpha t$) is available, we can verify that the reconstructed isovolume and the isosurfaces sliced from the isovolume are correct.

\textbf{Supernova simulation data.}\quad We reconstruct the isovolume of a supernova simulation dataset~\cite{Morozova18} with the isovalue of 0.8 and visualize the sliced isosurfaces of four timesteps in Figure~\ref{fig:isovolume}(b).  The resolution of the data is $432^3$, which in turn produces $\sim6$G spacetime edges to test intersections for the 4-timestep data.  As a result, the isovolume consists of 25M intersection points and 141M tetrahedra, which can be sliced back into individual timesteps for visualization and analysis.

\subsection{Comparison with existing isosurface-tracking algorithms}

Compared with existing methods of isosurfacing in higher dimensions~\cite{BhaniramkaWC04, JiSW03}, the benefits of our method include (1) straightforward implementation, (2) no disambiguation cases, (3) guaranteed robustness, and (4) straightforward parallelization with GPUs and distributed environments.  First, the two-pass algorithm can be written in a few lines of code based on FTK meshing APIs; there is no need to generate large lookup tables for higher-dimensional marching cubes.  Second, compared with marching cubes~\cite{LorensenC87}, no ambiguity cases exist with simplicial meshes.  Our method can be viewed as a generalization of marching tetrahedra~\cite{DoiK91} in higher dimensions, which automatically eliminates any ambiguities.  Third, the use of the simplicial mesh makes it possible to ensure robustness with symbolic perturbations, leading to consistent tracking results.  Fourth, our two-pass algorithm can be easily distributed and computed with parallel computing resources.

%% file: library.tex
\section{FTK library design}
\label{sec:library}

FTK's software design takes into account parallelization for both distributed- and shared-memory environments,  and the needs of both in situ and post hoc analyses.  

\textbf{Simplicial spacetime mesh APIs.}\quad 
FTK provides three functions, namely, \texttt{element\_for}, \texttt{sides}, and \texttt{side\_of}, to support the development of feature-tracking algorithms that traverse mesh elements along different dimensions of spacetime simplical meshes.  The \texttt{element\_for} function takes a user-defined labmda function as the input and enables the traversal of all $k$-simplices.  The \texttt{sides} function returns the set of $(k-1)$-simplicial sides for the given $k$-simplex; the \texttt{side\_of} function returns the set of $(k+1)$-simplices, whose side contains the input $k$-simplex.
All three functions are frequently used in the implementation of feature-tracking algorithms in FTK.  For example, in the first pass of 3D critical point tracking (Section~\ref{sec:critical}), we use \texttt{element\_for(3, detect\_critical\_point)} to detect critical points in all 3-simplicial cells in the 4D spacetime mesh.  In the second pass, we use both \texttt{sides} and \texttt{side\_of} to help determine how triangular faces should be connected.

\textbf{Inline numerical functions.}\quad
FTK implements numerical functions---small-matrix linear algebra and symbolic perturbations---for feature-tracking algorithms.  The implementation is header-only and template-based, such that the numerical functions can be directly compiled with C/C++/CUDA and executed in GPU kernel functions.  For example, in 3D critical point tracking, the robust critical point test relies on the sign of a determinant calculated by symbolic perturbation.  If test positive, the exact location of the critical point can be estimated by solving a linear system, and the type of the critical point is determined by the eigenvalues of the Jacobian matrix.

\textbf{Distributed union-find.}\quad
We use an asynchronous distributed union-find method~\cite{xu2020distributed} to enable distributed feature tracking.
Union-find is a key algorithm for partitioning a set of mesh elements into disjoint subsets.  
By eliminating frequent and expensive global synchronizations, this method outperforms existing implementations based on the bulk-synchronous parallel programming model.  

\textbf{I/O.}\quad FTK can be built with various I/O libraries, including VTK~\cite{vtk}, NetCDF~\cite{RewD90}, HDF5~\cite{hdf5}, and ADIOS2~\cite{LofsteadZKS09}, to read and write data in a variety of formats.  Data streamed in situ can be loaded into memory with ADIOS2.  FTK's output formats include VTK, text, and Python objects.  With VTK, trajectories and surfaces are transformed into \texttt{vtkPolyData}; isovolumes are written in \texttt{vtkUnstructuredGrid} formats.  FTK also supports human-readable text formats.  Python objects can be retrieved and serialized into either JSON or \texttt{pickle} formats.


%% file: interface.tex
\section{In situ and post hoc analysis with FTK}
\label{sec:interface}

FTK provides four different utilities for end users: VTK/ParaView filters, a command line interface, Python bindings, and a C++ programming interface.  The VTK/ParaView filters are designed for interactive visualization, with the possibility to couple simulations through ParaView Catalyst and other in situ frameworks.  The command line interface can be used for loosely coupled in situ processing and post hoc analyses.  The Python bindings are designed for post hoc analyses and integration with data science libraries.  The C++ programming interface is designed mainly for tightly coupled in situ analyses.  Readers can find documents and examples in the FTK repository: \url{https://github.com/hguo/ftk}.

\begin{figure}
  \includegraphics[width=\linewidth]{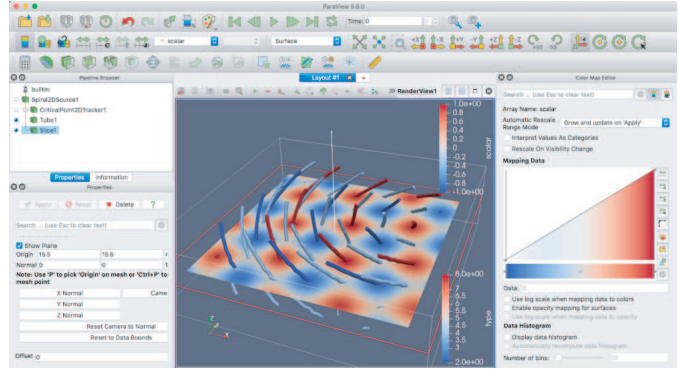}
  \caption{Example use of FTK's ParaView plugins.}
  \label{fig:paraview}
\end{figure}

\textbf{ParaView Plugins.}\quad We developed ParaView data sources and filters for users to use FTK interactively.  Synthetic data sources, including double gyre, spiral woven, and merger functions demonstrated in this paper, are available for users to learn FTK filters.
FTK filters include \texttt{vtkCriticalPointTracker} and \texttt{vtkLevelsetTracker}.  Currently, the inputs need to be image data types, and the output \texttt{vtkPolyData} can be directly rendered and further processed with other filters with ParaView.  Figure~\ref{fig:paraview} demonstrates a critical point tracking case: a scalar field is generated from a synthetic data source, and then critical point trajectories are extracted and transformed into tubes for visualization.

\textbf{Command line interface.}\quad
We provide executables for tracking critical points, quantum vortices, and isosurfaces in data obtained from files and in situ data streams.  For file inputs, FTK supports multiple file formats, including NetCDF, HDF5, VTK, ADIOS2, and raw binaries.  Users need to explicitly specify variable names for self-described formats.  For in situ data streams, users must specify ADIOS2 stream sources, variable names, mesh information, and other needed parameters.  The command line interface automatically loads and handles data in parallel if executed with MPI.  Users are also provided with options to use multithreading and GPU accelerators.  

\textbf{Python bindings} enables post hoc feature tracking and analysis in Python.  PyFTK functions take \texttt{NumPy} arrays as inputs, allowing users to load data with Python's \texttt{netCDF4}, \texttt{h5py}, and other I/O libraries.  PyFTK also enables easy integration with other libraries. 
One may load scalar field data with the NetCDF4 Python module, apply Gaussian smoothing in spacetime with \texttt{SciPy}, and then track critical points with our Python bindings.  

\textbf{C++ programming interface.}\quad C++ APIs enable users to couple simulation codes with FTK and/or customize feature-tracking algorithms.  The former can be achieved with user-level APIs; the latter requires the use of developer-level APIs.  
User-level APIs are available to users to feed time-varying data into FTK algorithms.  Users can call \texttt{push\_field\_data()} functions to feed individual timesteps to FTK.  Developer-level APIs are provided to customize feature-tracking algorithms by using meshing APIs to access spacetime mesh elements.

%% file: performance.tex
\section{Performance Evaluation}
\label{sec:performance} 

\begin{figure}
  \includegraphics[width=\linewidth]{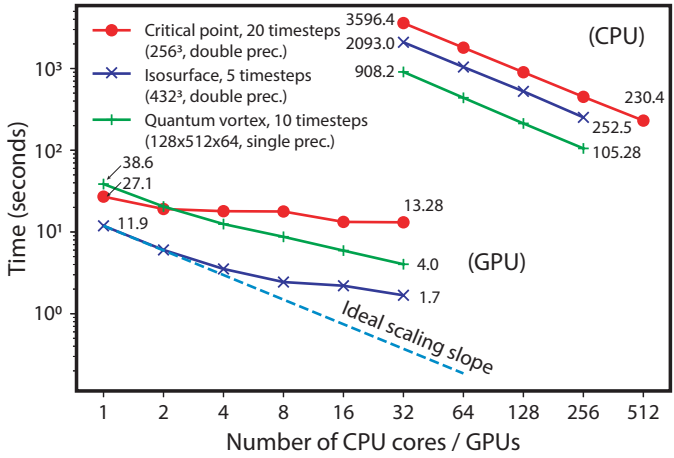}
  \caption{Timings, in seconds, of critical point tracking, isovolume, and quantum vortices computations for different numbers of CPU cores and GPUs on the Summit supercomputer.}
  \label{fig:benchmark} 
\end{figure}

We benchmarked the scalability of FTK on Summit, a 200-petaflop supercomputer at Oak Ridge National Laboratory.  The system consists of 4,608 IBM AC922 computing nodes, each of which has two 22-core Power9 CPUs and six NVIDIA Tesla V100 GPUs.  The clock frequency of each CPU is 3.07 GHz, and each node is equipped with 1,600 GB of high-bandwidth memory.  The interconnection between nodes is 100 Gbps EDR InfiniBand.  As shown in Figure~\ref{fig:benchmark}, we characterize the strong scalability by measuring the execution time of solving three feature-tracking problems with different numbers of processes, each of which uses either one GPU or one CPU core. 

\textbf{GPU acceleration.}\quad Compared with the CPU performance, the GPU acceleration typically ranges from $O(100)$ to $O(1000)$.  The acceleration is expected because there is no synchronization or  communication between GPU threads; each thread independently tests a mesh element.  The magnitude of GPU acceleration varies, possibly because of different complexities of handling mesh elements, precision of numeric algorithms (double vs. single), and data access cost (e.g., accessing scalars, vectors, and Jacobians in critical point tracking vs.\ accessing only scalars in isosurface tracking). We will study performance variability in future work.  

\textbf{Scalability.}\quad CPUs appear to scale better than GPUs.  As we continue to increase the number of processes each by a factor of 2, the acceleration of using GPUs saturates faster than when using CPUs.  Because the iteration over simplices is accelerated by GPUs, the non-GPU cost (e.g., data movement and connected component labeling) dominates when the workload per process decreases. In our applications such as fusion and superconductivity simulations, which typically produce 3 to 4 timesteps per second~\cite{CODAR2020, GuoPG17}, our algorithms are able to keep up with the data-producing rate in situ.  In the future we will further investigate the performance and resource consumption during in situ processing.

%% file: discussion.tex
\section{Discussion}
\label{sec:discussion}

We discuss FTK's design limitations, lessons learned, and comparison with other visualization libraries.

\subsection{Design limitations}

FTK builds on top of SoS and simplicial spacetime subdivisions and thus inherits limitations of both techniques.

\textbf{Limitations of SoS.}\quad SoS produces consistent results regardless of degeneracies given the predetermined vertex ordering system, but different vertex ordering systems may lead to different results because of SoS. 
In handling degeneracies, certain choices are implicitly made to associate features with mesh elements.  We will investigate how different vertex ordering could change the results in the future.

\textbf{Limitations of simplicial subdivision.}\quad Simplicial subdivision comes at a price although it has advantages in producing robust and combinatorial tracking results.  As documented by Carr et al.~\cite{CarrMS06}, subdivisions of cubic cells could lead to visual artifacts and topology changes in isosurfaces based on the choice of subdivision schemes.  With simplicial spacetime subdivision, we observe similar artifacts in critical point trajectories and isovolumes in the presence of noise.  In future work, we will investigate how different spacetime subdivisions affect feature-tracking results, and we will offer different triangulation schemes in FTK.

\subsection{Lessons learned from developing FTK}

Our objectives---simplifying, scaling, and delivering feature-tracking algorithms---align with the design goals of existing frameworks.  We share the filter and pipeline design patterns with other frameworks, which enable easy extension and integration in the future.

\textbf{Simplified time-varying data access.}\quad A key advantage of FTK is treating space and time dimensions equally in implementing feature-tracking algorithms.  For example, in the reconstruction of critical point trajectories in 4D spacetime, the tracking algorithm requires  access to the 4D volume, but time-varying data are usually read/produced in a streaming manner.  FTK's design allows hiding the details of time-varying data access by distinguishing ordinal/interval elements in the spacetime meshing design.  If one needs to implement a new algorithm similar to Algorithm~\ref{alg:tracking}, the out-of-core time-varying data access and parallelism are already managed by our framework.

\textbf{Synthetic data and unit tests.}\quad We evaluate synthetic data to verify our feature-tracking algorithms with ground truth.  We tailor synthetic data in order to incorporate corner cases rarely seen in real data.  For example, in the moving extremum case, we design critical point trajectories that intersect vertices and edges, in order to test whether the output trajectory is one single line with the expected initial location and direction.  As one of many unit tests in FTK, we use random numbers to define expected trajectories (Figure~\ref{fig:nonrobust}) and to verify the correctness of the outputs using single/multiple MPI processes, with or without GPUs.

\subsection{Comparison with other visualization libraries}

\textbf{Comparison with general-purpose visualization libraries.}\quad Compared with VTK~\cite{vtk} and VTK-m~\cite{MorelandSULMPKS16}, FTK features the spacetime meshing design and eases the access to time-varying data.  We hide the details of accessing/caching 4D data with the \texttt{element\_for()} function; a developer can use the same lambda function to iterate both ordinal and interval mesh elements in 4D.  To the best of our knowledge, the majority of VTK data structures and filters are purposed for individual timesteps; one needs to manage and cache time-varying data if a filter needs the access to consecutive timesteps in a visualization pipeline.

\textbf{Comparison with scalar field topology libraries.}\quad Compared with TTK~\cite{TiernyFLGM18}, FTK enables the tracking of a different family of features---``local'' features that can be localized in individual mesh elements---as opposed to ``global'' features such as merge trees, Morse-smale complexes, and Jacobi sets.  For the tracking of local features, FTK enables data parallelism and out-of-core data management because there are no task dependencies in mesh element traversal.

%% file: conclusions.tex
\section{Conclusions and Future Work}

This paper demonstrates the use of spacetime simplicial meshes for simplifying, scaling, and delivering feature-tracking algorithms, which are implemented in a suite of feature-tracking tools called FTK.
In the future we plan to investigate several aspects.  First, we will expand FTK's support of features including parallel vectors (1D features in 3D), ridge/valley surfaces (2D features in 3D), and interval volumes (3D features in 3D).  Second, we will develop spacetime mesh generalizations for cubic cells, finite volume meshes, and adaptive mesh refinements for broader scientific applications.  Third, we will incorporate scales as the fifth dimension for scale-space tracking.  